\documentclass[aps,prc,twocolumn,showpacs,bibnotes,footinbib,preprintnumbers,superscriptaddress,amsmath,amssymb]{revtex4-2}
\usepackage[T1]{fontenc}
\usepackage[utf8]{inputenc}
\usepackage{graphicx}
\usepackage{epsfig}
\usepackage{bm}
\usepackage{color}
\usepackage{float}
\usepackage{dcolumn}
\usepackage{txfonts}
\usepackage{multirow}

\begin{document}

\title{Decay study of $^{11}$Be with an Optical TPC detector}

%
\author{ N. Sokołowska}
\affiliation{Faculty of Physics, University of Warsaw, 02-093 Warszawa, Poland}
\author{ V.~Guadilla}
\affiliation{Faculty of Physics, University of Warsaw, 02-093 Warszawa, Poland}
\author{ C.~Mazzocchi}
\affiliation{Faculty of Physics, University of Warsaw, 02-093 Warszawa, Poland}
\author{R.~Ahmed}
\affiliation{National Centre for Physics, Islamabad, Pakistan}
\author{ M.~Borge}
\affiliation{Instituto de Estructura de la Materia, CSIC, Serrano 113-bis, E-28006 Madrid, Spain}
\author{ G.~Cardella}
\affiliation{INFN-Sezione di Catania, Catania, Italy}
\author{ A.A.~Ciemny}
\affiliation{Faculty of Physics, University of Warsaw, 02-093 Warszawa, Poland}
\author{L.G.~Cosentino}
\affiliation{INFN-Laboratori Nazionali del Sud, Catania, Italy}
\author{E.~De Filippo}
\affiliation{INFN-Sezione di Catania, Catania, Italy}
\author{V.~Fedosseev}
\affiliation{EN department, CERN, 1211 Geneva 23, Switzerland}
%
%
\author{ A.~Fijałkowska}
\affiliation{Faculty of Physics, University of Warsaw, 02-093 Warszawa, Poland}
\author{ L.M.~Fraile}
\affiliation{Grupo de F\'{\i}sica Nuclear \& IPARCOS, Universidad Complutense de Madrid, CEI Moncloa, E-28040 Madrid, Spain}
\author{E.~Geraci}
\affiliation{Dipartimento di Fisica e Astronomia “Ettore Majorana”, Università di Catania, Catania, Italy}
\affiliation{INFN-Sezione di Catania, Catania, Italy}
\author{ A.~Giska}
\affiliation{Faculty of Physics, University of Warsaw, 02-093 Warszawa, Poland}
\author{B.~Gnoffo}
\affiliation{Dipartimento di Fisica e Astronomia “Ettore Majorana”, Università di Catania, Catania, Italy}
\affiliation{INFN-Sezione di Catania, Catania, Italy}
\author{C.~Granados}
\affiliation{EN department, CERN, 1211 Geneva 23, Switzerland}
\author{ Z.~Janas}
\affiliation{Faculty of Physics, University of Warsaw, 02-093 Warszawa, Poland}
\author{ Ł.~Janiak}
\affiliation{Faculty of Physics, University of Warsaw, 02-093 Warszawa, Poland}
\affiliation{National Centre for Nuclear Research, PL 05-400 Świerk-Otwock, Poland}
\author{K.~Johnston}
\affiliation{ISOLDE, CERN, 1211 Geneva 23, Switzerland}
\author{ G.~Kamiński}
\affiliation{Heavy Ion Laboratory, University of Warsaw, 02-093 Warszawa, Poland}
\author{ A.~Korgul}
\affiliation{Faculty of Physics, University of Warsaw, 02-093 Warszawa, Poland}
\author{ A.~Kubiela}
\affiliation{Faculty of Physics, University of Warsaw, 02-093 Warszawa, Poland}
\author{C.~Maiolino}
\affiliation{INFN-Laboratori Nazionali del Sud, Catania, Italy}
\author{B.~Marsh}
\affiliation{EN department, CERN, 1211 Geneva 23, Switzerland}
\author{N.S.~Martorana}
\affiliation{INFN-Sezione di Catania, Catania, Italy}
\author{ K.~Miernik}
\affiliation{Faculty of Physics, University of Warsaw, 02-093 Warszawa, Poland}
\author{P.~Molkanov}
\affiliation{Institute covered by a cooperation agreement with CERN}
\author{ J.~D. Ovejas}
\affiliation{Instituto de Estructura de la Materia, CSIC, Serrano 113-bis, E-28006 Madrid, Spain}
\author{E.V.~Pagano}
\affiliation{INFN-Laboratori Nazionali del Sud, Catania, Italy}
\author{S.~Pirrone}
\affiliation{INFN-Sezione di Catania, Catania, Italy}
\author{ M.~Pomorski}
\affiliation{Faculty of Physics, University of Warsaw, 02-093 Warszawa, Poland}
\author{A.M.~Quynh}
\affiliation{Nuclear Research Institute, 670000 Dalat, Vietnam}
\author{K.~Riisager}
\affiliation{Department of Physics and Astronomy, Aarhus University, DK-8000 Aarhus C, Denmark}
\author{A.~Russo}
\affiliation{INFN-Laboratori Nazionali del Sud, Catania, Italy}
\author{P.~Russotto}
\affiliation{INFN-Laboratori Nazionali del Sud, Catania, Italy}
\author{A. Świercz}
\affiliation{AGH University of Science and Technology, Faculty of Physics and Applied Computer Science, 30-059 Krakow, Poland}
\author{S.~Vi\~{n}als}
\affiliation{Instituto de Estructura de la Materia, CSIC, Serrano 113-bis, E-28006 Madrid, Spain}
\author{S.~Wilkins}
\affiliation{EN department, CERN, 1211 Geneva 23, Switzerland}
\author{ M.~Pf\"{u}tzner}
\email{pfutzner@fuw.edu.pl}
\affiliation{Faculty of Physics, University of Warsaw, 02-093 Warszawa, Poland}
%
%

\collaboration{ISOLDE Collaboration}
\noaffiliation
\date{\today}

\begin{abstract}
The $\beta$ decay of one-neutron halo nucleus $^{11}$Be was investigated using the Warsaw Optical Time Projection Chamber (OTPC)
detector to measure $\beta$-delayed charged particles. The results of two experiments are
reported. In the first one, carried out in LNS Catania, the absolute branching ratio for
$\beta$-delayed $\alpha$ emission was measured by counting incoming $^{11}$Be ions
stopped in the detector and the observed decays with the emission of $\alpha$ particle.
The result of 3.27(46)\% is in good agreement with the literature value.
In the second experiment, performed at the HIE-ISOLDE facility at CERN, bunches containing
several hundreds of $^{11}$Be ions were implanted into the OTPC detector followed by the detection of decays with the emission of charged particles. The energy spectrum of
$\beta$-delayed $\alpha$ particles was determined in the full energy range. It was analysed
in the R-matrix framework and was found to be consistent with the
literature. The best description of the spectrum was obtained
assuming that the two $3/2^+$ and one $1/2^+$ states in $^{11}$B are involved in the transition.
The search for $\beta$-delayed emission of protons was undertaken.
Only the upper limit for the branching ratio for this process of
$(2.2 \pm 0.6_{\rm stat} \pm 0.6_{\rm sys}) \times 10^{-6}$
could be determined. This value is in conflict  with the result published in
[Ayyad et al. Phys. Rev. Lett. 123, 082501 (2019)] but does agree with the limit reported in
[Riisager et al., Eur. Phys. J. A (2020) 56:100]
\end{abstract}


\maketitle

\section{Introduction}
\label{sec:intro}

The isotope of beryllium $^{11}_{~4}$Be has attracted the attention of both
experimental and theoretical physicists for a long time due to
a few interesting features. Contrary to the standard shell-model
picture, its ground-state spin-parity is $1/2^+$ instead of $1/2^-$ \cite{Deutsch:1968,Geithner:1999}.
This spin inversion, which occurs due to residual interactions and provides
an example of the disappearance of the $N=8$ magic number, was spotted already
in 1960 \cite{Talmi:1960}. Effectively, it results from a migration of the $\nu 2s_{1/2}$
orbital below the $\nu 1p_{1/2}$ one. This, together with a low ground-state
binding energy of 502~keV, is responsible for the single-neutron halo
character of $^{11}$Be \cite{Fukuda:1991,Geithner:1999}. A further consequence
is the unusually long half-life of $^{11}$Be, $T_{1/2}=13.76(7)$~s \cite{Kelley:2012}.
The full and detailed understanding of the parity inversion in this nucleus has been
achieved only recently within \emph{ab initio} theory by taking into account
continuum effects and three-nucleon forces \cite{Calci:2016}.

\begin{figure*}
 \includegraphics[width =1.3 \columnwidth]{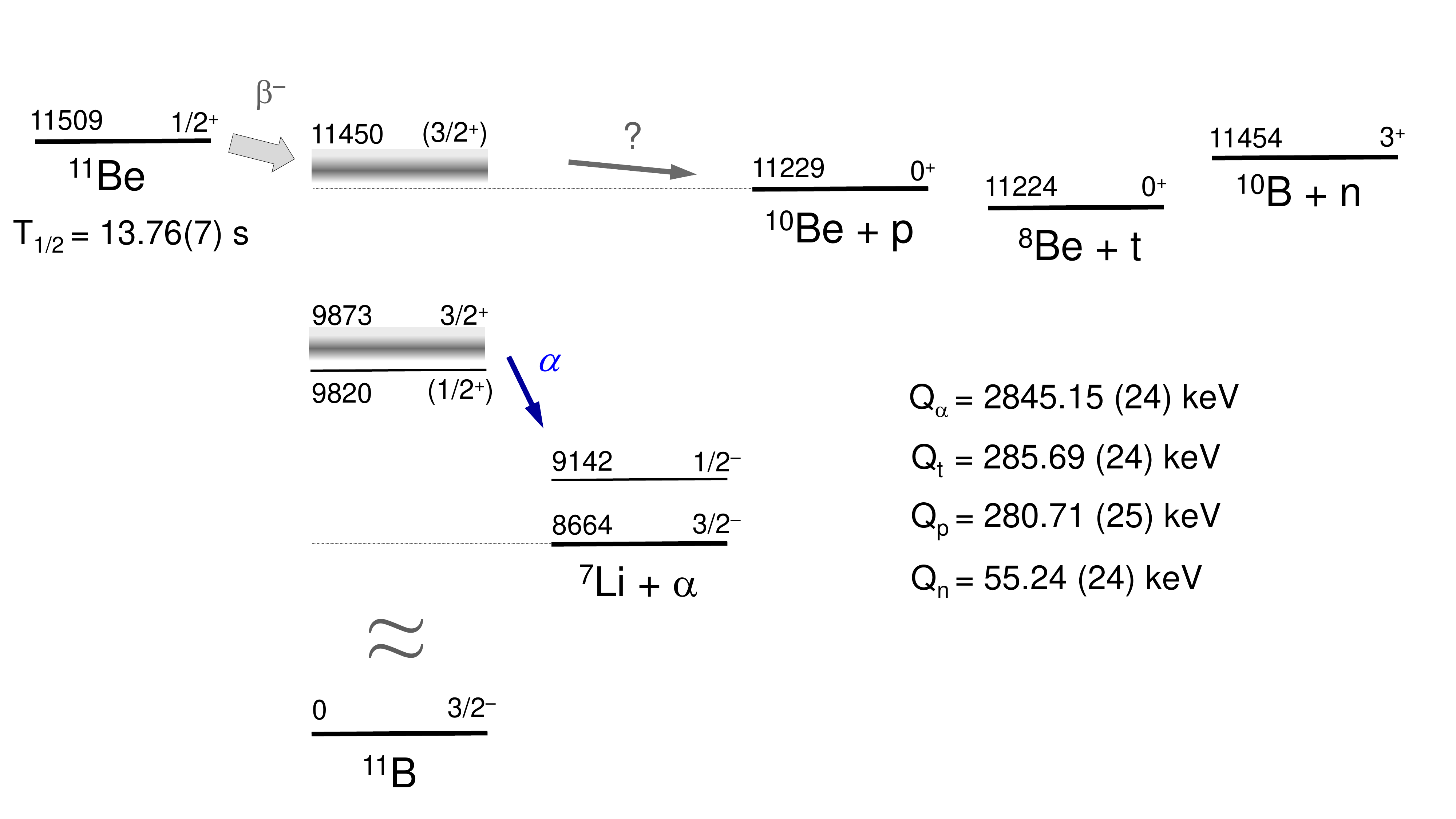}{}%
 \caption{Schematic decay scheme of $^{11}$Be. Level energies
  (in keV) are given with respect to the ground state of $^{11}$B. Only excited states
  in $^{11}$B which are above the particle emission thresholds, and can be fed by allowed
  $\beta$ decay, are shown, according to Ref.~\cite{Kelley:2012}. The $Q$-values
  for decays with emission of delayed particles were calculated using the
  AME 2020 mass evaluation tables \cite{Wang:2021}. }
  \label{fig:DecayScheme}
\end{figure*}

A part of the $^{11}$Be decay scheme, which is of interest for the present work,
is shown in Fig.~\ref{fig:DecayScheme}. The delayed emission of $\alpha$ particles ($\beta \alpha$)
was observed already in 1971~\cite{Alburger:1971}, but characterised with more details
in Ref.~\cite{Alburger:1981}, where the branching
ratio for the $\beta \alpha$ channel was measured to be $2.9(4)$\% and the
energy spectrum of $\alpha$ particles was explained by a single $\beta$ transition to
the $3/2^+$ state at 9.87~MeV in $^{11}$B. A more precise and recent
measurement of $\beta \alpha$ channel in the decay of $^{11}$Be was published by
Refsgaard et al.~\cite{Refsgaard:2019}. The value of the $\beta \alpha$ branching
ratio was determined to be $b_{\beta \alpha}=3.30(10)$\%. It confirmed
that the $3/2^+$ state at 9.87~MeV dominates the $\beta \alpha$ spectrum, however,
a better agreement with the data was obtained by inclusion of a second $3/2^+$ state
at 11.49(10)~MeV in $^{11}$B.

In the decay of $^{11}$Be also other delayed-particle channels are open, see
Fig.~\ref{fig:DecayScheme}. For the delayed emission of protons, tritons, and neutrons,
the available energy windows are 281~keV, 286~keV, and 55~keV, respectively.
The delayed emission of protons ($\beta p$) is of particular interest.
It is a very well known process among neutron-deficient nuclei and a valuable
source of nuclear-structure information for nuclei far from stability \cite{Pfutzner:2012}.
There are only a few neutron-rich nuclei, however, where this process
can occur \cite{Baye:2011}. From the energy considerations it follows that the
delayed emission of a proton after $\beta^-$ decay is possible only when the
available energy, $Q_{\beta p}=782 {\rm \, keV} - S_n$, is positive, where $S_n$ in the
neutron separation energy in the decaying nucleus \cite{Jonson:2001}. Thus, low
values of $S_n$ are required, which makes the neutron halo nuclei the prime
candidates \cite{Baye:2011}. In turn, the study of $\beta p$ emission provides
a tool to investigate the halo structure of the initial nucleus \cite{Riisager:2017,Borge:2013}.
The $^{11}$Be is one of these candidates. The early theoretical prediction for
its $\beta p$ branching ratio in a two-body potential model yielded the value
$b_{\beta p}=3.0 \times 10^{-8}$~\cite{Baye:2011}.

The experimental search for $\beta p$ emission from $^{11}$Be started 10 years
ago at ISOLDE/CERN \cite{Borge:2013}. Due to the expected very small probability of this
decay channel, a hybrid method was adopted. First, a sample of $^{11}$Be was
collected using ISOLDE mass separator. Then, the presence of $^{10}$Be - the $\beta p$
daughter of $^{11}$Be - was searched for in the sample by means of accelerator
mass spectrometry (AMS). The very first attempt did not yield a positive evidence
with the branching $b_{\beta p}=(2.5 \pm 2.5) \times 10^{-6}$~\cite{Borge:2013}.
The second approach, however, with the $^{11}$Be source collection at ISOLDE and
the AMS measurements made at the VERA facility at the University of Vienna, provided
an unexpectedly high value of $(8.3 \pm 0.9) \times 10^{-6}$ \cite{Riisager:2014}.

The interest in the direct observation of $\beta$-delayed protons from the decay of
$^{11}$Be was suddenly boosted when the hypothesis of a dark decay channel of the
neutron was put forward to explain discrepancies between the neutron lifetime
measurements \cite{Fornal:2018}. It was followed by an observation, that such
a dark neutron decay could occur also in some nuclei with $^{11}$Be being the most
promising candidate \cite{Pfutzner:2018}. Such a decay channel of this nucleus would
lead to $^{10}$Be, as claimed to be observed in Ref.~\cite{Riisager:2014}, however,
with no emission of protons. Soon afterwards, the direct emission of protons following
the decay of $^{11}$Be was reported by Ayyad et al.~\cite{Ayyad:2019} who employed the
Active Target Time Projection Chamber (AT-TPC) detector in an experiment performed at
the ISAC-TRIUMF laboratory. The branching ratio was found to be $(1.3 \pm 0.3) \times 10^{-5}$,
in agreement with the results of Ref.~\cite{Riisager:2014}. The observed energy
distribution of protons indicated that the decay proceeds through a narrow
resonance in $^{11}$B at $11425(20)$~keV, with a total width $\Gamma = 12(5)$~keV and
$J^{\pi}=(1/2^+, 3/2^+$)~\cite{Ayyad:2019}. These findings triggered a number of
theoretical attempts to interpret such a cluster-like, narrow resonance close
to the decay threshold \cite{Okolowicz:2020,Volya:2020,Elkamhavy:2021,Atkinson:2022,Okolowicz:2022,LeAnh:2022,Elkamhavy:2023}.
The situation became less clear, though, when authors of Ref.~\cite{Riisager:2014}
carried out another hybrid-like measurement in an attempt to reproduce their previous result.
After special efforts to produce clean samples of $^{11}$Be and a careful examination
of potential contamination sources, Riisager et al. \cite{Riisager:2020} concluded
that the formation of $^{10}$BeH$^+$ molecular ions in the ISOLDE ion source
was a probable source of contamination in the previous experiment, leading to the
surprisingly large $b_{\beta p}$ value. From the new measurement, only an upper
limit for the $\beta p$ branching ratio of $2.2 \times 10^{-6}$ could be
extracted \cite{Riisager:2020}. On the other hand, new evidence for a narrow,
near-threshold, proton-emitting resonance in $^{11}$B, consistent with the
results of Ref.~\cite{Ayyad:2019} came from reaction studies \cite{Ayyad:2022,Lopez-Saavedra:2022}.

It appears evident, that further independent experimental studies are needed
to verify the existence of the $\beta p$ decay branch of $^{11}$Be and, if confirmed,
to clarify the puzzle of its strength. Guided by this motivation, and encouraged by
a successful study of $\beta d$ emission from $^{6}$He \cite{Pfutzner:2015}, we undertook a
study of $^{11}$Be $\beta$ decay using the Warsaw Optical TPC (OTPC) detector to record
tracks of emitted charged particles. In contrast to the AT-TPC chamber used in
Ref.~\cite{Ayyad:2019}, our detector was operated with a gas mixture of higher
density which allowed us to observe the $\beta \alpha$ decay channel in the
full energy range, from 200 keV to 3 MeV. Two experiments were performed.
In the first one, at the INFN-LNS laboratory in Catania, the separated and identified in-flight
ions of $^{11}$Be were implanted into the OTPC one-at-a-time, and the following decays
with emission of $\alpha$ particles were observed. This allowed us to
determine independently the absolute branching ratio for the $\beta \alpha$ channel.
The second experiment was made at HIE-ISOLDE at CERN. Reaccelerated bunches containing
a large, but unknown number of $^{11}$Be ions were implanted into the
OTPC followed by observation of their decays. A large number
of $\beta \alpha$ events allowed to establish the energy spectrum for this channel
with a different method than the one used by Refsgaard et al.~\cite{Refsgaard:2019}.
This provided a check of the energy calibration of our chamber and, together with
the $\beta \alpha$ branching ratio, the absolute normalisation for the number
of implanted $^{11}$Be ions. The large statistics accumulated in this measurement
was sufficient to look for the $\beta p$ decay channel at the level of $10^{-6}$.
In this paper, we present the results of both experiments.

\section{Experimental techniques}

\subsection{The OTPC detector}
\label{sec:OTPC}

\begin{figure}
 \includegraphics[width = \columnwidth]{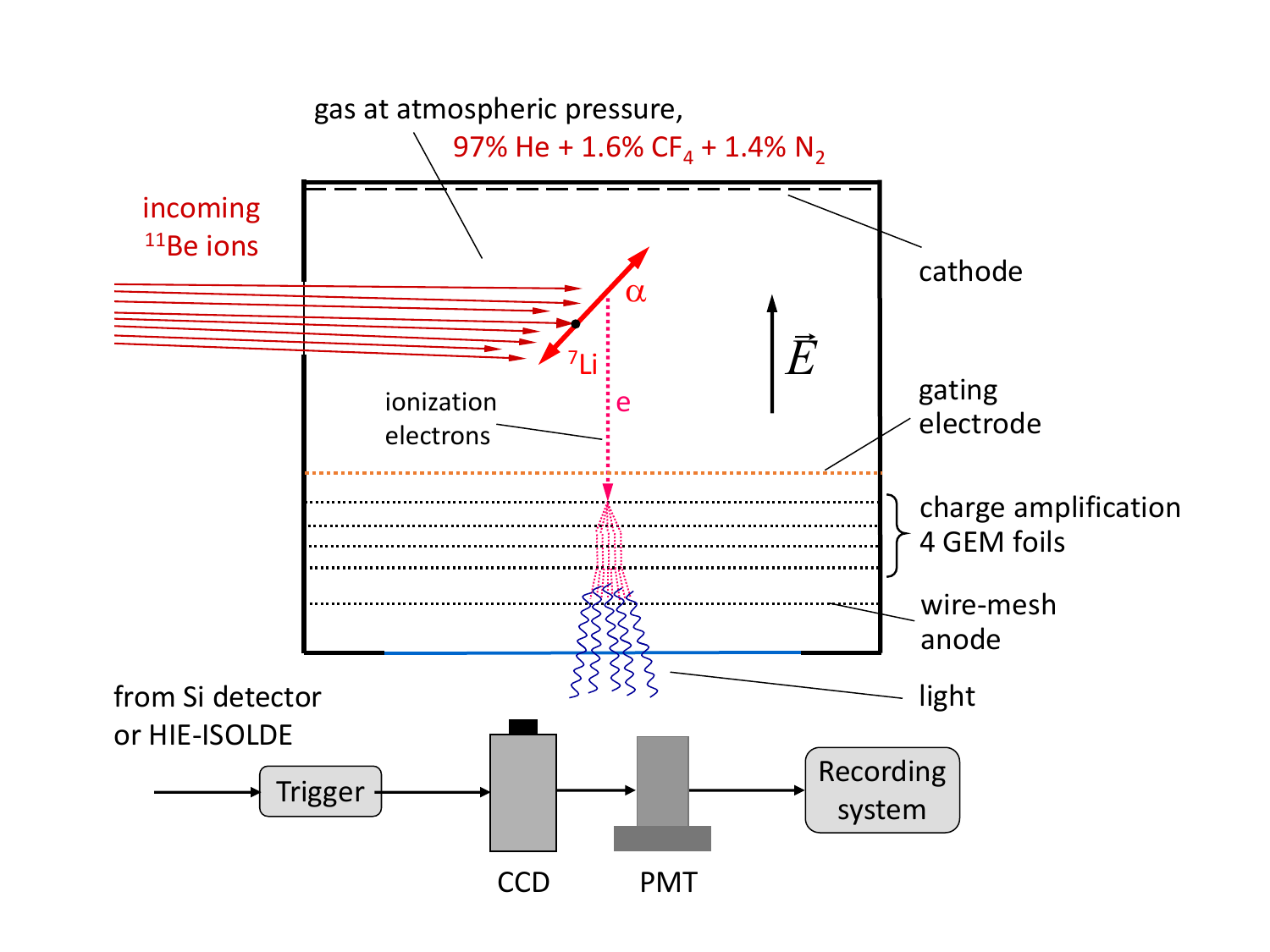}
 \caption{Schematic representation of the OTPC detector and its working principle. }
 \label{fig:OTPC_scheme}
\end{figure}

The OTPC detector was developed at the University of Warsaw
to study rare decay modes with emission of charged particles, like 2\emph{p} radioactivity
and $\beta$-delayed multi-particle emission. More details on this instrument are given
in Ref.~\cite{Pomorski:2014} and the newest, upgraded version is described
in Ref.~\cite{Ciemny:2022}. Here we summarize briefly the principle of operation with
a focus on features relevant to this work. The scheme of the detector is shown in
Fig.~\ref{fig:OTPC_scheme}.

In the reported experiments, the detector was filled with a gaseous
mixture of 97\% He, 1.6\% CF$_4$, and 1.4\% N$_2$
at atmospheric pressure.
Within the active volume, between the cathode and the amplification stage, a
constant and uniform electric field of 143 V/cm in the vertical direction was maintained.
Ions enter the active volume horizontally, through a kapton window.
Primary ionization electrons, generated by the stopping ion and by charged particles
emitted in its decay, drift with a constant velocity $\upsilon _d$ towards the
amplification stage, passing first through a wire-mesh gating electrode.
By changing the potential of this electrode the sensitivity of the detector
can be modified. This feature is used to reduce or block the large charge
generated by heavy ions. The charge amplification is realized by
four gas electron multiplier (GEM) foils \cite{Sauli:1997}.
Below the GEM section, there is the final wire-mesh anode. In the space between
the last GEM foil and the anode, the electric signal is converted to light.
The light is recorded with a CCD camera and a photomultiplier (PMT)
connected to a digital oscilloscope with the 50~MHz sampling frequency.
The CCD image represents a projection of an event on the anode plane,
integrated over exposure time, while the PMT waveform provides
the total light intensity as a function of time. The latter
contains the information on the event along the direction of the
electric field, i.e. perpendicular to the anode plane. The combination
of data from the CCD and the PMT allows the reconstruction of the decay event
in three dimensions, provided no particle escaped the active volume.
In front of the OTPC entrance window, an aluminium degrader of variable
thickness is mounted to optimize the implantation depth of the ions
of interest. In addition, during the experiment at the LNS, a Si detector
of 140 $\mu$m thickness was placed in front of this degrader.
It provided an additional energy-loss signal, $\Delta E$, used to
confirm the particle identification.

The gas flowing out of the OTPC was passing through another small gas
chamber where the electron drift velocity was measured. The same value
of the electric field was maintained to ensure the same conditions as in the
OTPC. The average value of the $\upsilon _d$ was found to be 9.20~mm/$\mu$s and
9.36~mm/$\mu$s for the LNS and ISOLDE experiments, respectively.

\subsection{Experiment at LNS}
\label{sec:ExpCatania}

\begin{figure}
 \includegraphics[width = \columnwidth]{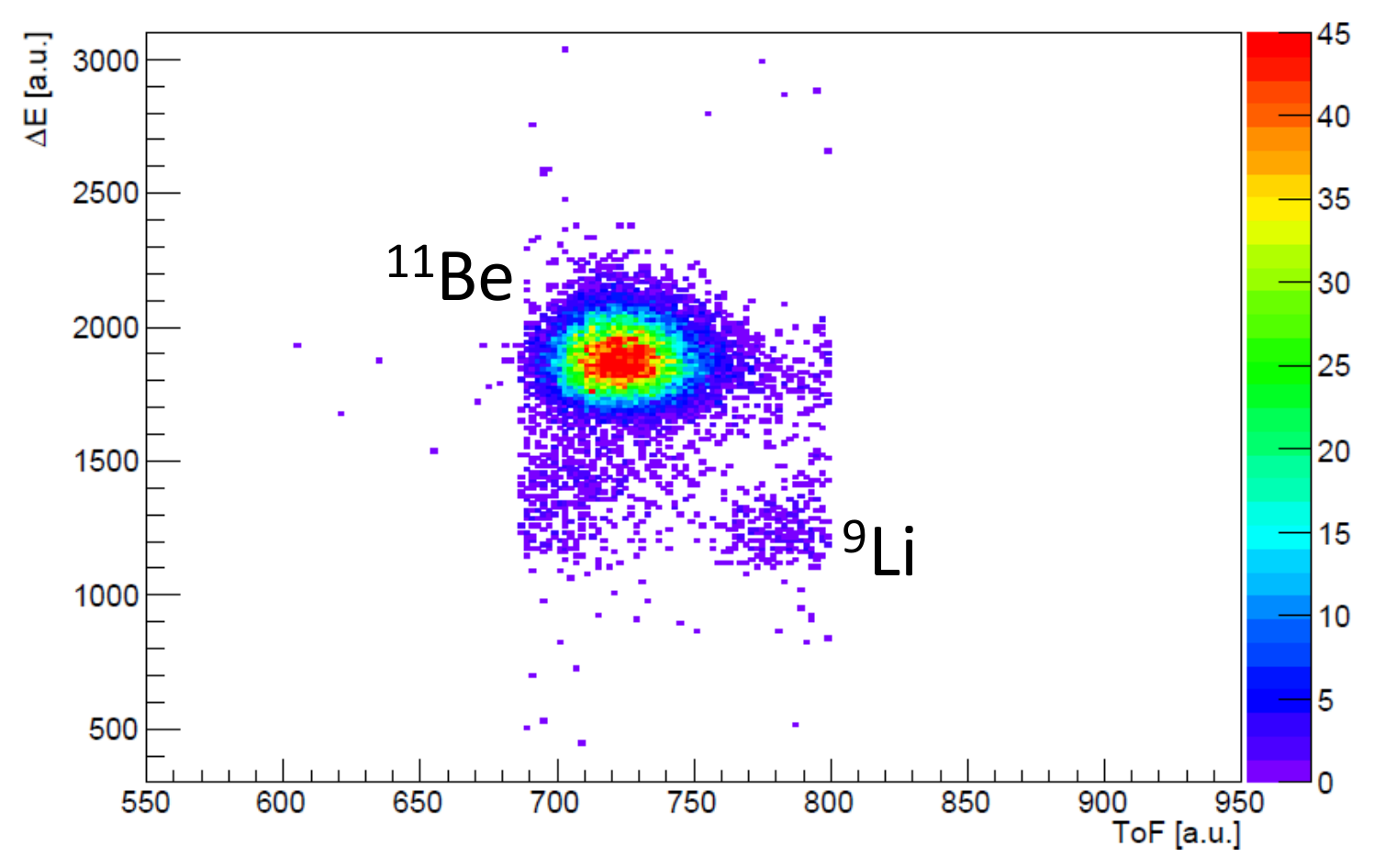}
 \caption{(Color online) The $\Delta E$--TOF identification
 plot of ions delivered by the separator from the FRIBs acquisition system.
 Contaminant ions of $^{9}$Li  are seen on both sides of the measured TOF band
 due to periodic condition implied by using the RF signal for the time reference.}
 \label{fig:ID_Catania}
\end{figure}

The measurement was performed in the Laboratori Nazionali del Sud of INFN (INFN-LNS)
in Catania, Italy, at the in-Flight Radioactive Ion Beams (FRIBs)
facility \cite{Russotto:2018,Martorana:2022}.
Ions of $^{11}$Be were produced by the fragmentation reaction of a $^{13}$C
primary beam, delivered by the Superconducting Cyclotron (SC), impinging
on a 1.5~mm thick beryllium target at the energy of 55~MeV/nucleon.
The reaction products were purified in a separator composed of
two 45$^{\circ}$ dipole magnets and a homogenous aluminium 1~mm thick degrader
mounted between them. The main feature of this production method,
of key importance for this work, is the possibility of full identification
in-flight of single ions coming out of the separator. This was accomplished
by the $\Delta E$--TOF technique. The energy-loss ($\Delta E$) information was
provided by a 70~$\mu$m thick DSSSD detector, mounted at
the entrance to the so-called 0$^{\circ}$ experimental hall,
where the OTPC detector was installed.
The time-of-flight (TOF) measurement was started by the Radio
Frequency (RF) signal from the SC and stopped by signals from the DSSSD strips.
The resulting identification plot is shown in Fig.~\ref{fig:ID_Catania}.
The main group of events represents ions of $^{11}$Be, while
small contamination comes from ions of $^{9}$Li.
The average rate of $^{11}$Be was about 200 ions/s.

\begin{figure}
 \includegraphics[width = \columnwidth]{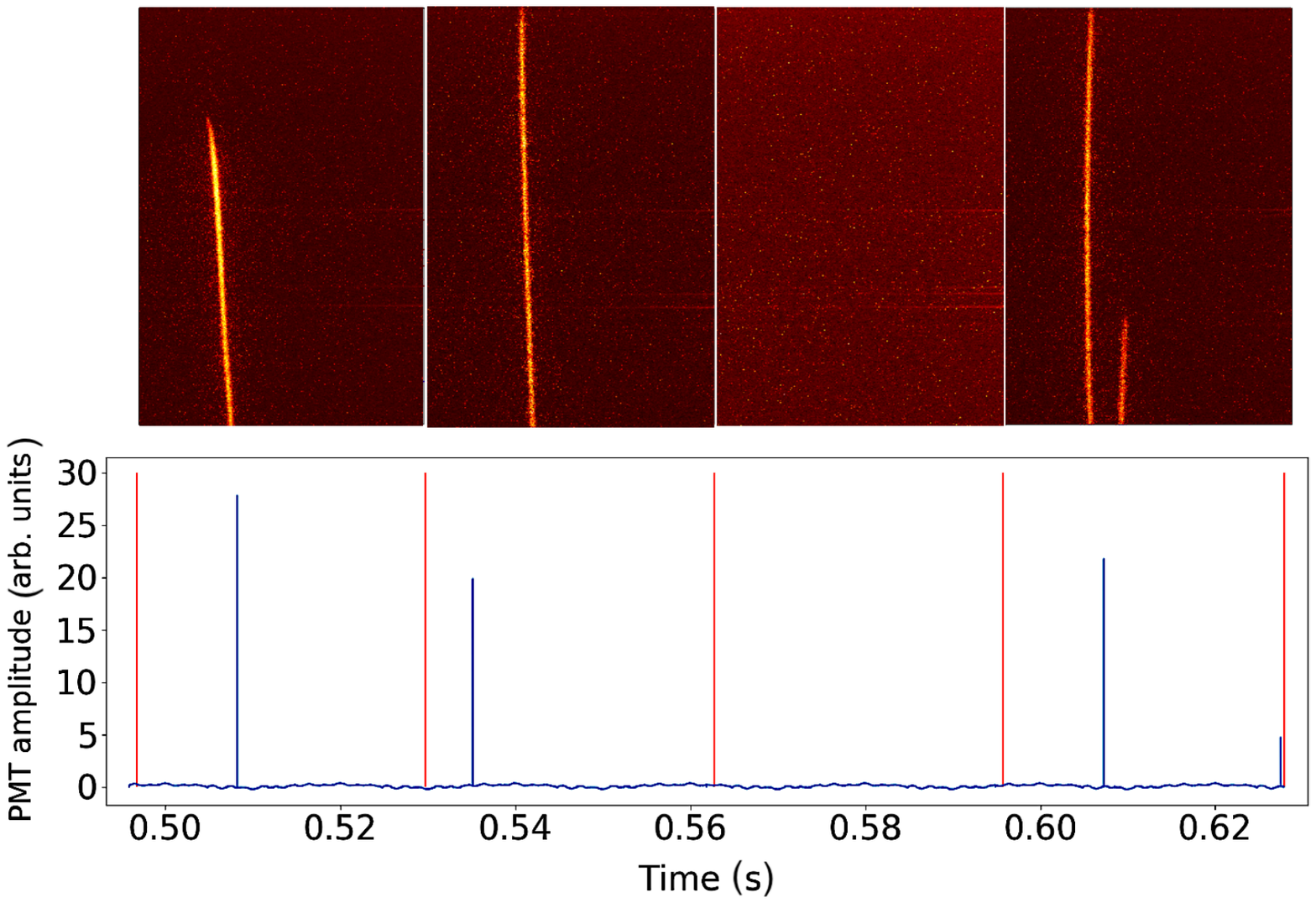}
 \caption{(Color online) Example of data taken at the LNS. The first few CCD images (top)
 with the corresponding PMT waveform (bottom) showing the ions entering the OTPC
 detector. Time values separating individual CCD frames are marked on the
 PMT waveform with red lines. In the first and the fourth frame, a stopped
 ion can be seen.}
 \label{fig:Catania_Ions}
\end{figure}

The main goal of the experiment at the LNS was to remeasure the branching
ratio $b_{\beta \alpha}$ for the $\beta \alpha$ emission from $^{11}$Be.
Since identified in-flight single ions were being implanted into the
OTPC, the $b_{\beta \alpha}$ could be determined by
counting the number of stopped ions and the number of decays with the emission
of $\alpha$ particle. Due to a large momentum spread of $^{11}$Be ions,
being a consequence of the fragmentation reaction in a relatively thick
target, the range distribution of these ions was broader than the
gas thickness of the OTPC. To determine the fraction of $^{11}$Be ions
stopped in the active gas section of the detector,
a dedicated measurement was made, where ions, implanted into the OTPC in
an uninterrupted way, were counted.
The CCD was operated in a cycle of taking 27 images, each with the 33~ms
exposure, followed by a 5~s break to save the data on the disk.
The PMT was recording the light continuously over the time of all 27 images.
An example showing the beginning of such a ``movie'' is given in Fig.~\ref{fig:Catania_Ions}.
In this part of the experiment more than 100 thousand CCD images were collected.

\begin{figure}
 \includegraphics[width = \columnwidth]{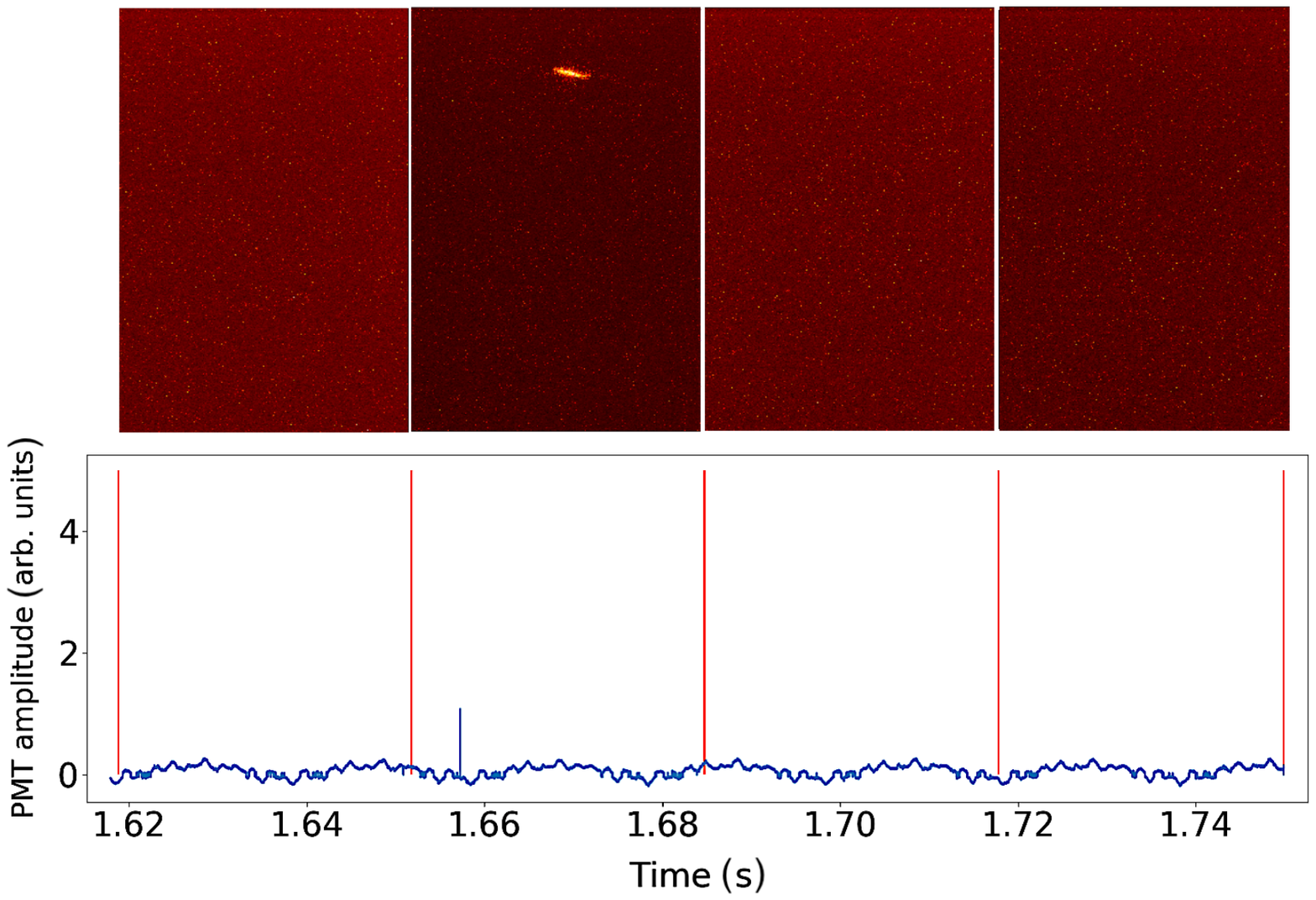}
 \caption{(Color online) Example of date taken at the LNS. The first few CCD images (top)
 with the corresponding PMT waveform (bottom) showing the decays of $^{11}$Be
 ions stopped during the preceding beam-on period in the OTPC
 detector. Time values separating individual CCD frames are marked on the
 PMT waveform with red lines. Only in the second frame, a decay event
 is observed}
 \label{fig:Catania_Decays}
\end{figure}

A different detection cycle was used in the runs where $\beta \alpha$ decay
events of $^{11}$Be were counted. Using a special low-energy chopper, at the
ion source of the SC, the beam was delivered to the OTPC only in short periods
of 750~ms, every 60~s.
After such implantation period, the OTPC acquisition system was started. A series of
63 CCD images, of 33~ms exposure each, was recorded, while the PMT was recording
light continuously over the time of all these images. This set  was followed
by a 1.2 s break to save the data on the disk. Such a sequence was repeated 8 times.
Thus, the collection of the decay data
lasted about 26 seconds. In the remaining 34 seconds
of the beam cycle no data were recorded. This time was set to make sure
that most of the stopped ions will decay before the next bunch of ions is
implanted.
As an example, a part of the decay data is shown in Fig.~\ref{fig:Catania_Decays}.
In total, almost 4 million CDD images were collected in this part of the experiment.

\subsection{Experiment at HIE-ISOLDE}
\label{sec:ExpIsolde}

The $^{11}$Be beam was produced at the ISOLDE facility at CERN \cite{Borge:2017,Catherall:2017}
and post-accelerated at the HIE-ISOLDE accelerator \cite{Kadi:2017}.
The 1.4 GeV protons from the PS Booster accelerator were directed onto
a UCx target equipped with a tantalum hot cavity.
Beryllium isotopes were laser ionised using the RILIS ion source \cite{Fedosseev:2017}
and accelerated to 30~keV before being mass separated by the general-purpose mass separator.
For the beam to be post-accelerated, it was injected first into a Penning trap (REXTRAP),
where the beam was bunched, and then into an electron beam ion source (EBIS)
where the singly charged $^{11}$Be was charge-bred to $^{11}$Be$^{4+}$ with a breeding
time of 46~ms. The beam, with $A/q = 2.75$, was subsequently injected into a linear
post accelerator where an eventual energy of 7.5~MeV/nucleon was achieved.
To remove ions of $^{22}$Ne$^{8+}$, which were present as an impurity from the EBIS,
carbon stripping foils were employed allowing a very clean beam of $^{11}$Be
to be delivered, via the XT03 beam line, to the
experimental station where the OTPC detector was installed.

\begin{figure}
 \includegraphics[width = \columnwidth]{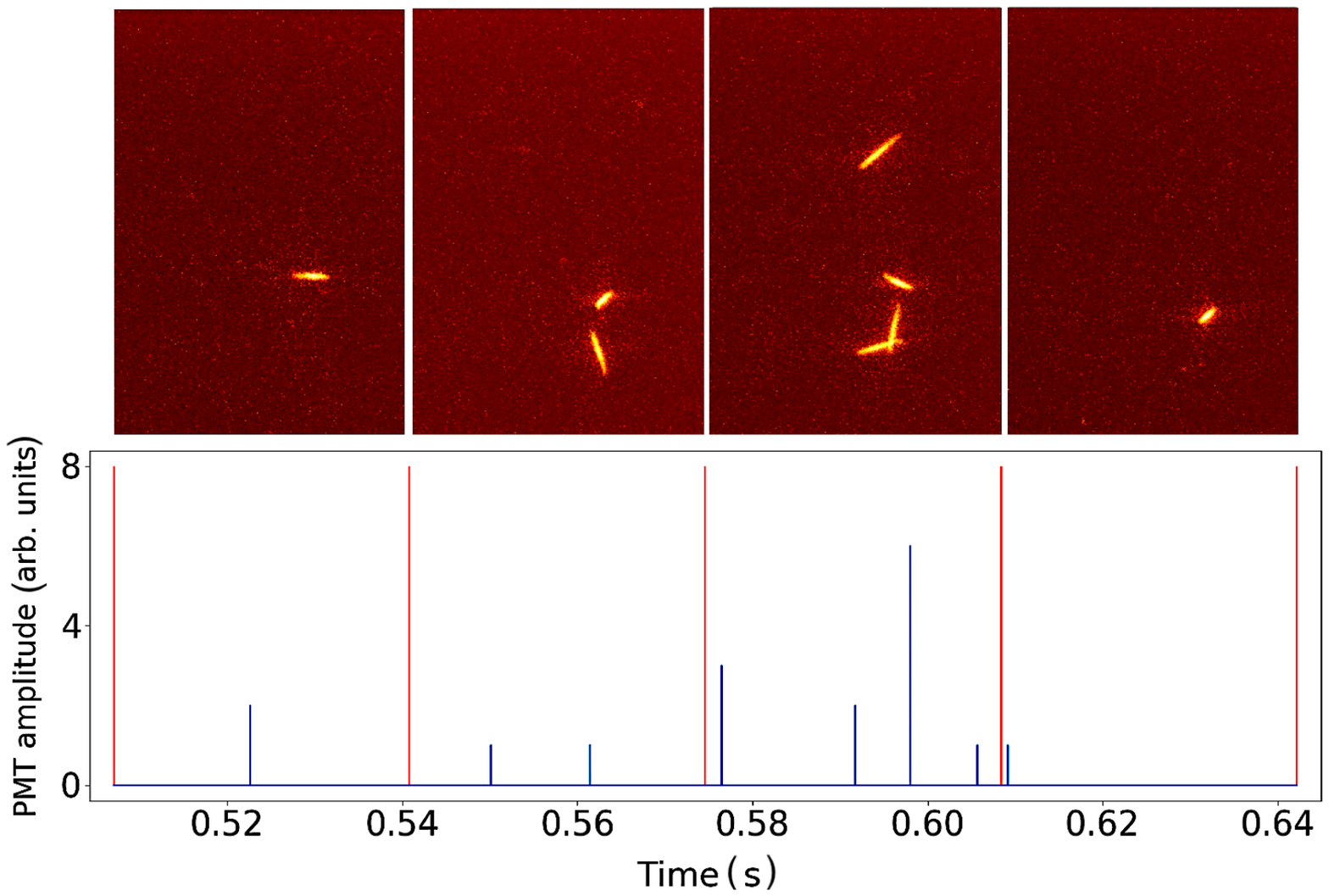}
 \caption{(Color online) Example of data taken at the HIE-ISOLDE. The first few CCD images (top)
 with the corresponding PMT waveform (bottom) showing decays of $^{11}$Be
 ions stopped during the preceding beam-on period in the OTPC
 detector. Time values separating individual CCD frames are marked on the
 PMT waveform with red lines.}
 \label{fig:Isolde_Decays}
\end{figure}

The detection cycle was similar to the one used at LNS for the
decay data collection. It started with the opening of the beam gate
for 750~ms during which $^{11}$Be ions were implanted. Afterwards,
the OTPC acquisition system was started.
A series of 63 CCD images, with 33~ms exposure, were
taken while the PMT was recording the light continuously during this time.
A break of 1.2 s followed, to save the data on disk. This sequence was
repeated 4 times, yielding a total detection time of about 12~s.
For the next 48~s, no data were recorded. After the 60~s from the start
the next bunch of ions was accepted and the new detection cycle was started.
An example of the collected data is shown in
Fig.~\ref{fig:Isolde_Decays}.

\section{Data analysis}
\subsection{Single-ion implantation}

Data collected in the LNS experiment allow us to determine the number
of $^{11}$Be ions stopped inside the detector and count the number of
decays with emission of charged particles. From these, the branching
ratio for $\beta$-delayed $\alpha$ emission can be calculated.

The first part of the LNS run was used to establish the percentage of
ions stopped in the OTPC. An automatic counting procedure was developed
for this purpose.
Each CCD image was divided into 10 horizontal slices along $x$-axis (perpendicular to the beam direction). The content of each slice was projected on the $x$-axis. An algorithm based on a peak searching routine was applied to each slice in order to perform tracking of the ion trajectories, allowing the determination of the end of the track for those ions stopped inside the chamber. The good performance of the routine was verified in a subset of data. The PMT waveform corresponding to the CCD image was used to cross-check the number of ion tracks found by the algorithm.
From the results obtained, the number of ions entering the OTPC and the
number of ions stopped within the active volume was determined.
After applying this procedure to all CCD frames in this part, the
final average probability of stopping an ion in the OTPC was
found to be 19.4(27)\%. Events with a mismatch between the number
of tracks in the CCD and the PMT data represent the main contribution to the error.
The distribution of the $y$ coordinate of the
stopped ions (range distribution in the OTPC gas) was found to be almost
flat, in agreement with the calculations made with the LISE$^{++}$
ion-optical simulation tool \cite{LISE:2016}.

The final identification of ions entering the OTPC detector was
made with the help of the Si detector mounted in front of the
chamber. The ID plot obtained with this detector was essentially
the same as the one provided by the DSSSD detector used by the
standard FRIBs acquisition system, shown in Fig.~\ref{fig:ID_Catania}.
Since some ions could have been stopped in materials between the Si
detector and the active gas volume (variable Al degrader, entrance window),
the counting of ions entering the OTPC was made by applying
a coincidence condition between the signal from the Si detector and
the light detected in the OTPC by the PMT. From the resulting ID plot,
it was found that $^{11}$Be ions represented 93(1)\% of ions entering
the OTPC. The remaining 7\% corresponded to ions of $^{9}$Li.

In the second part of the LNS experiment, the information on $^{11}$Be
decays was recorded. From the ID data collected during this part, we
counted 633 501 ions, which entered the OTPC detector.
After the correction for beam purity and stopping probability,
about 114300 ions of $^{11}$Be were found as stopped inside the detector.
Most of the 4 million CCD images taken in this part were empty.
A selection of frames (about $10^5$) containing a clear signal above the noise
was chosen for further analysis.
A majority of them showed a track of $\alpha$ particle from a weak
diagnostic $\alpha$ source installed inside the OTPC,
composed of $^{243}$Cm, $^{243}$Am, and $^{241}$Am. These tracks
were easy to discern because they emerged from the well-defined
region on the left border of the image and their length
was much longer than the longest tracks expected in the decay of $^{11}$Be.
Another type of events originated from natural radioactive decay chains,
which included $\alpha$ emitters. They could be also distinguished from
the $^{11}$Be decays by the much larger energy, and thus the track length.
The third source of background was represented by the point-like, short
light flashes on the GEM foils. They were easy to identify due to the
small size of a few pixels on the CCD image and a very short signal of a
few samples in the PMT waveform. A dedicated selection tool was developed
which scanned all recorded frames and using the characteristics mentioned
above removed the frames containing these background events.
After this procedure, 1837 decay events remained. Among this number
some events may come from decays of the contaminant $^{9}$Li which
has a 50\% branch of $\beta$-delayed neutrons, leading to two $\alpha$
particles in the final state \cite{Prezado:2005}.
Since the half-life of $^{9}$Li is only 178~ms, decays of any
stopped $^{9}$Li will happen within the first sequence of 63 CCD
images (the first ``movie'') in the decay-detection cycle which
lasts 2.079~s (see Section~\ref{sec:ExpCatania}). To avoid this
contamination, we dismissed 380 decay events which were found to
occur in the first ``movie''. Because the branching
ratio for the delayed emission of protons in $^{11}$Be is more than
3 orders of magnitude smaller than that for the emission of $\alpha$
particles, we can safely assume that the resulting number of 1457 decay
events represents only $\beta \alpha$ emission from $^{11}$Be.
As the observation time of decays was finite,
the corresponding correction must be introduced. Taking into account the
detailed timing cycle of the decay measurement, with the first ``movie''
removed, and the half-life of $^{11}$Be, the probability of recording
an $\beta \alpha$ decay was 39.0(1)\%.

Taking together relevant numbers and correction factors, we arrive at
the branching ratio for $\beta$-delayed $\alpha$ emission,
$b_{\beta \alpha} = 3.27(46)$\%. It agrees very well with
the value of $3.30(10)$\%, reported in Ref.~\cite{Refsgaard:2019}, albeit is less
precise. The uncertainty of our value is dominated by the uncertainty
of the probability of stopping ions in the OTPC detector.

\subsection{Decay events}

The number of decay events collected during the LNS experiment was too
small for the analysis of the $\beta \alpha$ energy spectrum.
This goal can be achieved by the analysis of data from the HIE-ISOLDE
experiment where much larger statistics of $^{11}$Be decays was gathered.
In total about one million CCD images were taken but, in contrast to
the LNS experiment, many frames contained two or more decay events.
In these cases, there is no general unambiguous way to decide which
event seen on the CCD image corresponds to which decay signal in the
PMT waveform. That is why  we selected only frames exhibiting one decay
event for the further analysis. There were about 270 thousand such frames.

\begin{figure}
 \includegraphics[width = 0.8\columnwidth]{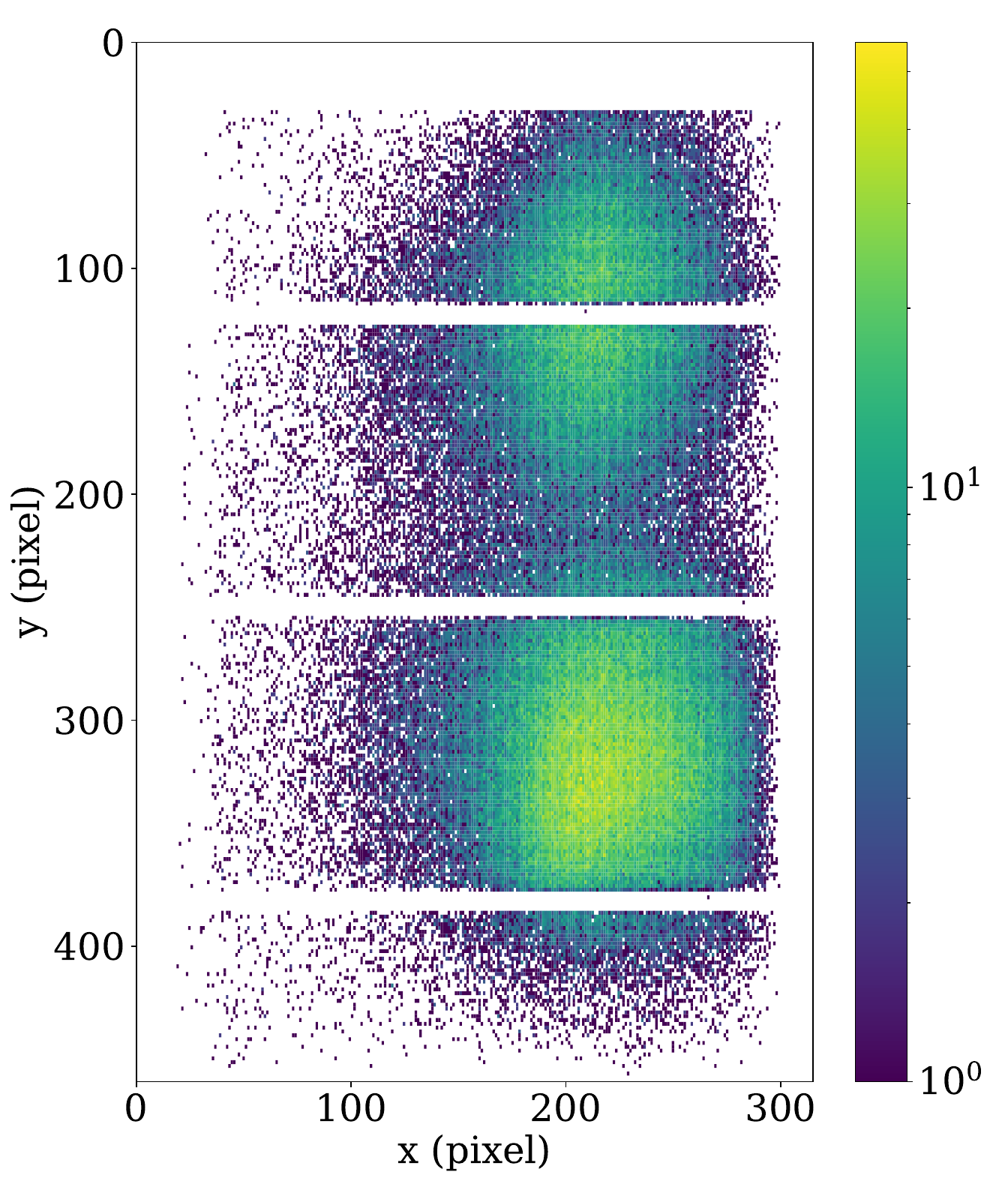}
 \caption{(Color online) Position distribution of selected $^{11}$Be decay
 events on the CDD plane in the HIE-ISOLDE experiment. One pixel corresponds
 to 0.63~mm. See text for more explanations. }
 \label{fig:DecayPositionIsolde}
\end{figure}

In the next step, each CCD frame was analysed by a procedure detecting
the position of the event on the image, and in particular determining
the positions of both ends.
All events, where the position of one of the ends of the track resulted
to be closer than 2 cm from the wall, were discarded.
This is because close to the walls the electric
field is less homogeneous and thus the reconstruction of such events is
less reliable. In addition, we have to take into account that the GEM
foils used in the OTPC are composed of four sections which are
separated electrically by a narrow inactive strip which stops drifting electrons.
As a consequence, on the image plane there are three horizontal narrow bands
with strongly reduced sensitivity. The events which were found to overlap with
one of these bands were also discarded. The resulting position distribution
of the decay-event centers, on the image plane ($x$, $y$), is shown in Fig.~\ref{fig:DecayPositionIsolde}. The decays can be seen to concentrate
in two separate regions which correspond to two locations in the detector
where the incoming ions were stopped. This is caused by the OTPC entrance
window which is covered with horizontal strips of 5~$\mu$m of copper and
2~$\mu$m of gold used to maintain the uniform electric field in
the chamber. The strips are 7~mm wide with 3~mm space in between.
Although the energy of the $^{11}$Be beam was well defined, the beam
spot on the OTPC window was broader than one strip, so that ions passing
through the additional strip material lose more energy, and thus have
a shorter range in the gas than ions passing between the strips.
We cannot determine the decay position in the vertical ($z$) direction,
so some decays can happen close to the cathode or the GEM section and
may appear deformed or damaged. Due to diffusion, the probability
of finding an ion far from the implantation point, in all directions,
increases with time and about 5\% of ions implanted in one detection
cycle can survive more than 1 minute and decay within the next cycle.
To partly moderate this effect, we decided to discard events having the position
coordinate $x < 100$~pixels, see Fig.~\ref{fig:DecayPositionIsolde}.

\begin{figure}
 \includegraphics[width = \columnwidth]{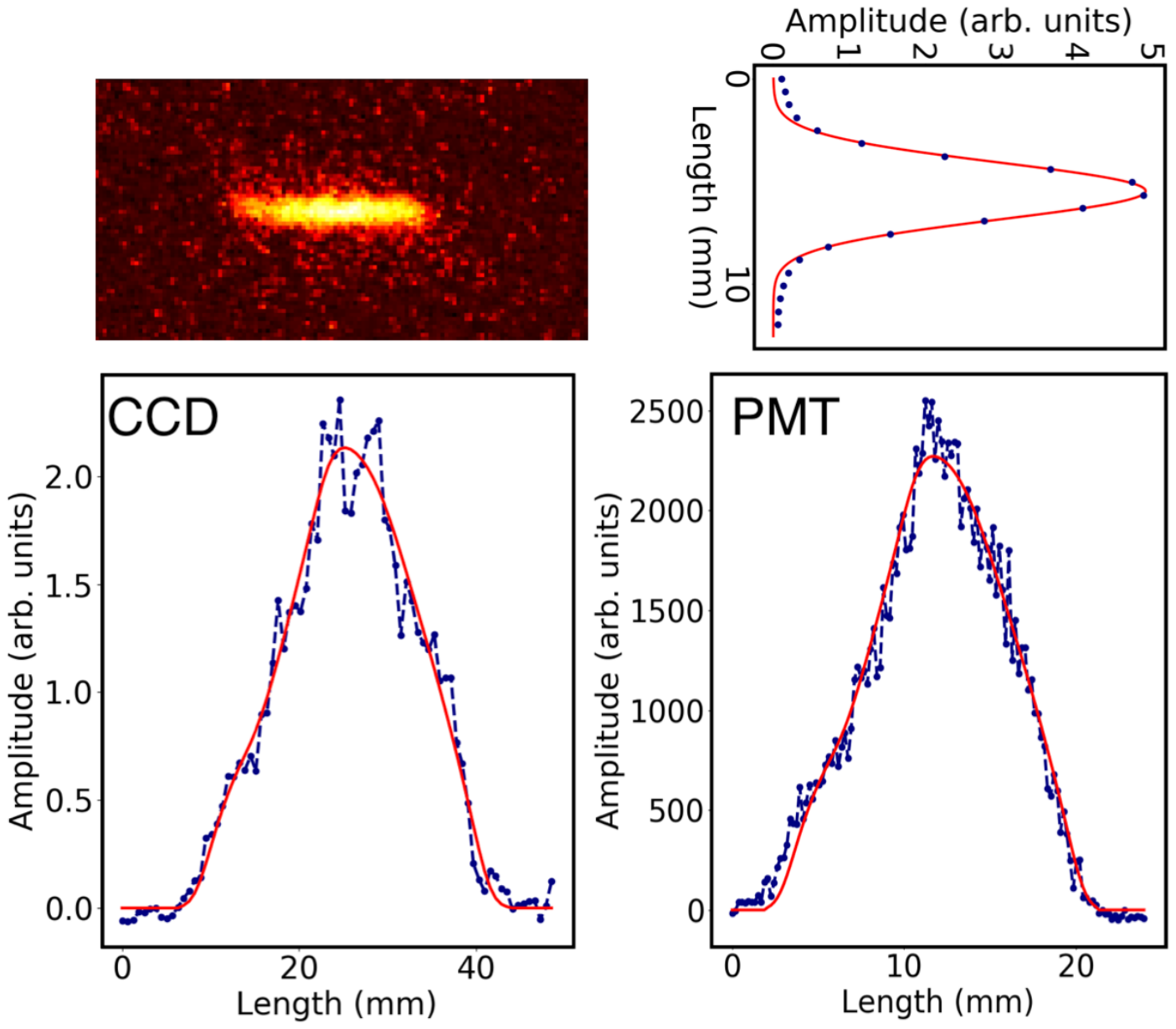}
 \caption{(Color online) Example data for one decay event of $^{11}$Be.
 Top left: a part of the CCD image showing the event track;
 top right: the projection of the image part on the axis perpendicular to the track with the best-fitted Gaussian curve (red line);
 bottom left: the projection of the image part on the axis parallel to the track;
 bottom right: the corresponding part of the PMT signal representing the vertical
 component in the length scale. The red lines in the bottom panels show the
 best-fitted model signal corresponding to a $\beta \alpha$ event with the decay
 energy of 1140~keV and the emission angle of the $\alpha$ particle of $28^{\circ}$
 with respect to the horizontal plane.     }
 \label{fig:DecayEvent}
\end{figure}

Since the $\beta$-delayed emission in $^{11}$Be happens almost at rest,
the particles are emitted in opposite directions (back-to-back)
forming a single, straight track, as can be seen in
Figs.~\ref{fig:Catania_Decays} and \ref{fig:Isolde_Decays}.
The reconstruction of each event in three dimensions is done by
extracting the ionisation distribution along this track and by
comparing it with the energy loss model. First,
the relevant parts containing the decay signal were isolated from the PMT
waveform and from the CCD image of each event. A rectangular part containing
the track was cut out from the image and its
content was projected, in the image plane, on the direction along the track,
and on the direction perpendicular to it.
The former projection was very well approximated by a Gaussian
curve with the width parameter $\sigma_{\rm{CCD}}$ describing all effects of
electron diffusion, in the drift and the amplification sections, in the horizontal
plane of the detector.
The latter projection represented the ionisation distribution
along the track as seen on the horizontal plane. An example
result for one decay event is shown in Fig.~\ref{fig:DecayEvent}.

In the analysis, we consider two decay scenarios: delayed emission
of $\alpha$ particle ($^{11}$B$^*\rightarrow \, ^{7}$Li$ + \alpha$)
and delayed emission of a proton ($^{11}$B$^*\rightarrow \, ^{10}$Be$ + p$).
The predictions for the ionization profile along the track
were done using two models for the energy loss of charged particles
in matter.

\begin{figure}
 \includegraphics[width = \columnwidth]{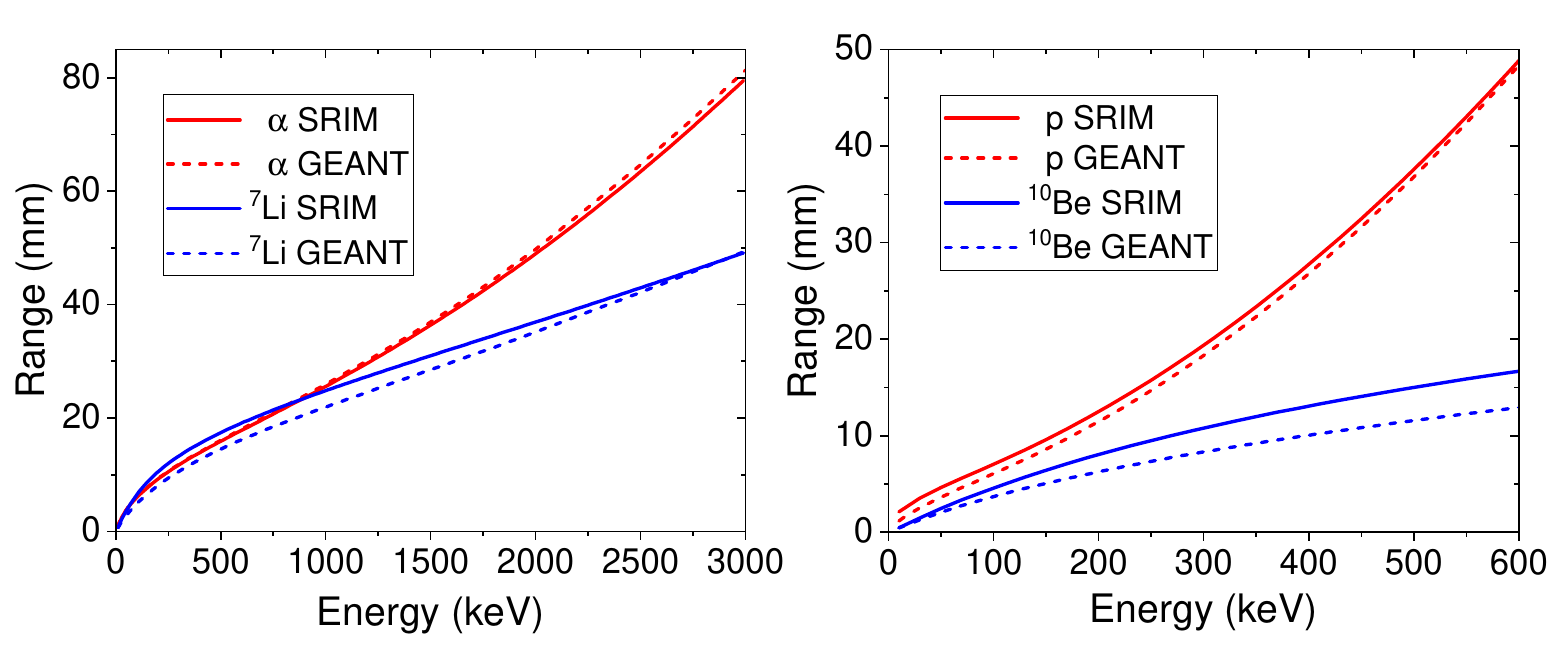}
 \caption{(Color online) Ranges of particles in the OTPC gas mixture used in the HIE-ISOLDE
 experiment, as a function of particle energy, as predicted by the SRIM \cite{SRIM:2010}
 and the GEANT4 \cite{GEANT4:2003} models.}
 \label{fig:Ranges}
\end{figure}

\begin{figure}
\includegraphics[width = \columnwidth]{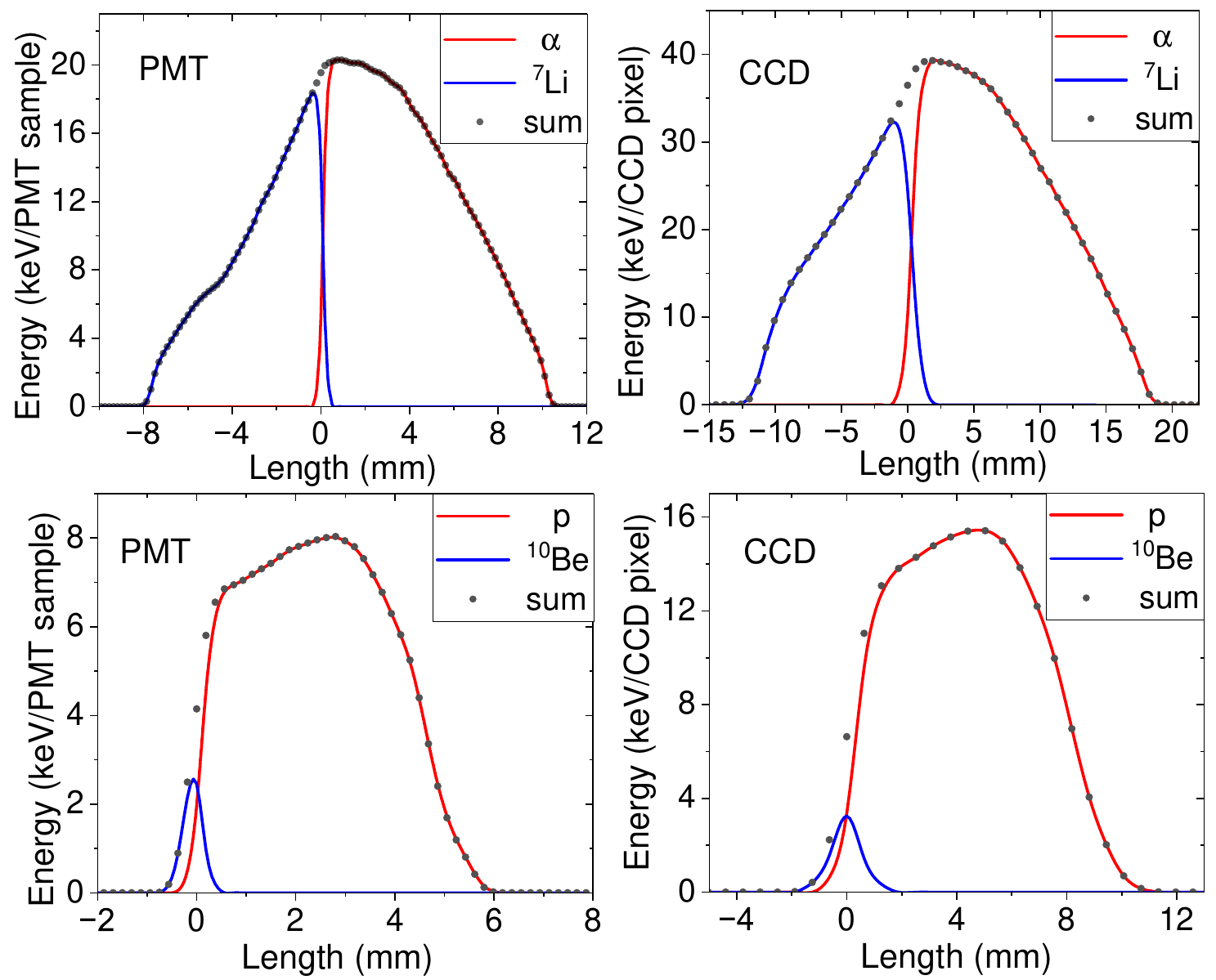}
 \caption{(Color online) Predicted energy deposition profiles calculated with the
 SRIM model for the decays of $^{11}$Be in the OTPC gas in the HIE-ISOLDE experiment.
 Top: decay into $\alpha + ^{7}$Li for the decay energy of 1200 keV and the $\alpha$ emission
 angle of $30^{\circ}$ with respect to the horizontal plane. Bottom: decay into $p + ^{10}$Be for
 the decay energy 200 keV and the proton emission angle of $30^{\circ}$. On the left the vertical
 projections are shown, as seen in the PMT signal, on the right the profiles along the
 track projected on the horizontal plane are shown, as extracted from the CCD image.
 All curves were diffused with $\sigma_{PMT} = \sigma_{CCD} =2$ samples/pixels.  }
 \label{fig:ModelEvents}
\end{figure}

The first one is the model included in the SRIM package \cite{SRIM:2010}, commonly adopted in similar applications. The second one is the low-energy model included in the GEANT4 simulation toolkit \cite{GEANT4:2003}, employing the ICRU49 parameterisation of evaluated data for stopping powers \cite{ICRU:1993}.
The use of the latter was necessary, as we used GEANT4 to make
realistic simulations of our data events to verify the reconstruction
procedure, as described below. With these models we first computed the ranges
of particles in the OTPC gas, as a function of their energy, see Fig.~\ref{fig:Ranges}.
Having these functions, one can calculate the energy deposited along the
track for the given decay energy. By the energy deposit we mean here the
energy lost by ionization of gas atoms - this caveat is important especially
at the end of the track, where the stopping power is dominated by nuclear
collisions which do not liberate electrons. Assuming the emission angle with respect
to the image plane ($x, y$), one can determine the expected distributions of
the energy deposit in the two measured projections of the track. In the final step,
before the comparison with the data,
these predicted distributions are folded with the Gaussian curves accounting
for the diffusion of the primary ionisation electrons during the drift time
and the spread due to electron multiplication in the GEM section.
For the horizontal projection, the $\sigma_{\rm{CCD}}$ value, extracted from the
image, was taken. For the vertical projection the corresponding $\sigma_{\rm{PMT}}$
value was considered as a free parameter in the minimization process.
Examples of total energy-loss model distributions
with segregated contributions from both particles are shown in Fig.~\ref{fig:ModelEvents}.
For each decay event, and for both decay scenarios, a least-squares minimization
procedure was applied to find the decay energy, emission angle with respect
to the horizontal ($x,y$) plane, and $\sigma_{\rm{PMT}}$
best describing the measured distributions. It was done by minimizing the
chi-square function:
\begin{equation} \label{eq:chi2}
\chi^2 = \sum_{i=1}^{n}\left(\frac{d_{\rm{CCD}}^{\, i} - m_{\rm{CCD}}^{\, i}}{\delta_{\rm{CCD}}}\right)^2 + \sum_{i=1}^{m}\left(\frac{d_{\rm{PMT}}^{\, i} - m_{\rm{PMT}}^{\, i}}{\delta_{\rm{PMT}}}\right)^2,
\end{equation}
where the first (second) sum runs over the $n$ ($m$) data points of the CCD (PMT) signal.
With $d_{c}^{\, i}$ and $m_{c}^{\, i}$ the values of the $i$-th data point and
the model are denoted, respectively, while $\delta_{c}$ represents the
uncertainty of the data in the channel $c$ (CCD or PMT).
For the uncertainties, the following approximation was adopted:
\begin{equation} \label{eq:DataErrors}
\delta_{c}^2 = \frac{\sum^{n}_{i=1}\left(d_{c}^{\, i} - d_{\rm{smooth}}^{\, i}\right)^2}{n},
\end{equation}
where the $d_{\rm{smooth}}$ is the result of smoothing the $d_c$ data set with
a Gaussian filter. An example result of this minimization procedure is shown in Fig.~\ref{fig:DecayEvent}.

To verify the reconstruction procedure described above, and to gauge its
performance, we used the GEANT4 package \cite{GEANT4:2003} to
make simulations of the observed decay events.
The real conditions in the OTPC were assumed
and the distribution of the energy deposited was simulated for a given
decay type, decay energy, and emission angle in the 3D space.
Then, the resulting distribution was diffused both in horizontal and vertical
directions using realistic widths. Finally, it was projected on the $(x,y)$
plane, and on the $z$ direction, taking into account the pixel size of
the CCD image (0.63~mm), and the binning of the PMT signal
(equal to $\upsilon _d \times 20$~ns), respectively.
A noise distribution, sampled from the experimental data, was added to
both output files. This procedure yields a CCD-like image file and a
PMT-like waveform, providing a realistic representation of a decay event,
for which the physical parameters are known. More details of the simulation
procedure are given in Ref.~\cite{Guadilla:2023}. The simulated events are
reconstructed in exactly the same way as the real data, using the above-mentioned
GEANT4 model of energy loss and the determined parameters can be compared to the input ones.

\begin{figure}
 \includegraphics[width = \columnwidth]{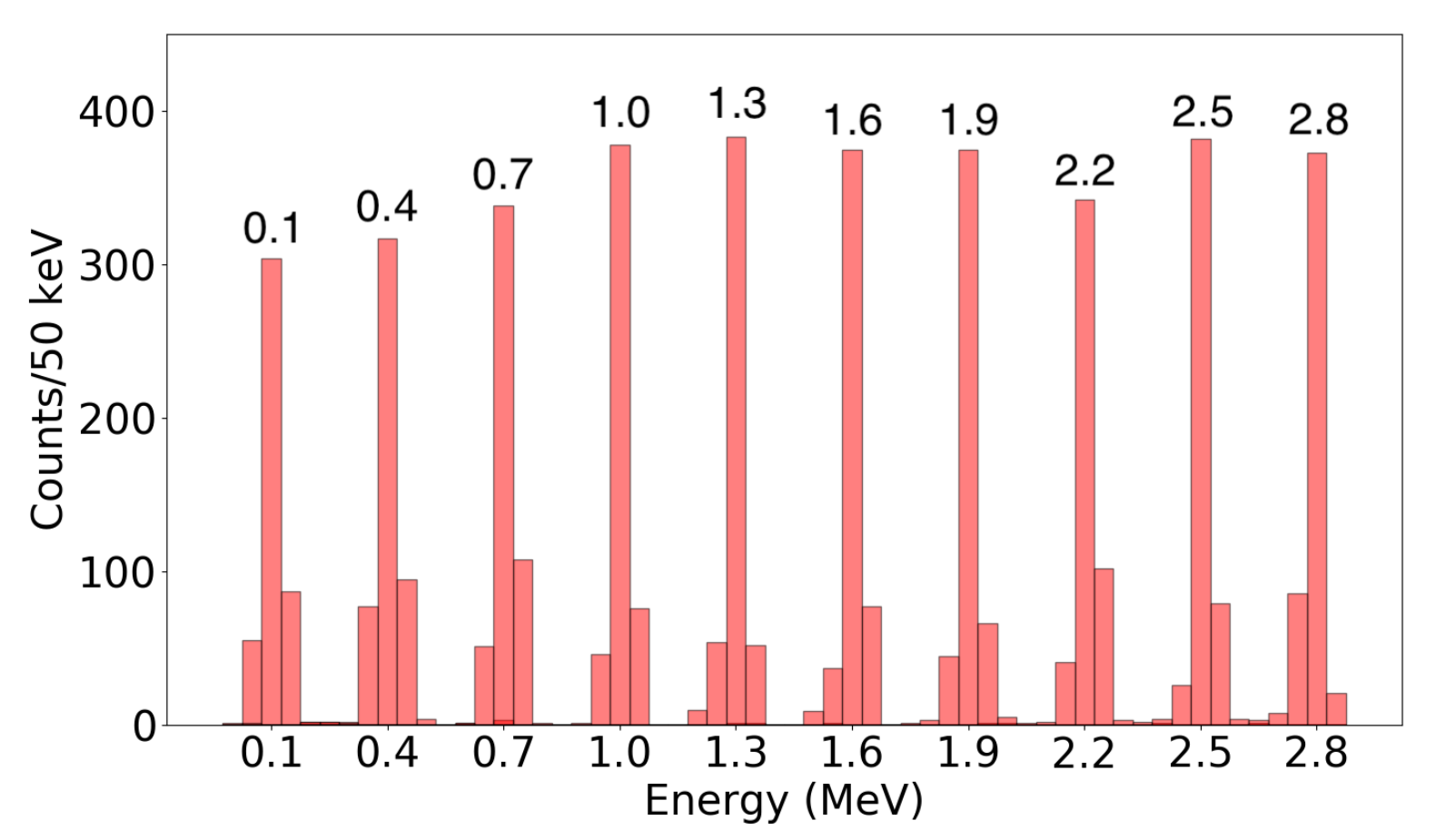}
 \caption{Results of reconstruction for a set of simulated
 monoenergetic $\beta \alpha$ decay events. Each peak corresponds to events of
 the same input energy shown by the label.}
 \label{fig:MonoSimulations}
\end{figure}

A set of simulations of the $^{11}$Be $\beta \alpha$ decay was made for energies in the
full energy range, assuming isotropic emission. In general, the reconstruction
procedure was found to reproduce the key input parameters very well.
In particular, the response of the reconstruction
to the simulated monoenergetic decays was studied systematically.
A set of $\beta \alpha$ events was simulated for well-defined energies in the
range from 100~keV to 2.8~MeV and for isotropic emission in space.
After reconstruction, each monoenergetic group was found in a Gaussian-like
distribution with its maximum equal to the input energy and the width ($\sigma$)
of about 25~keV, slightly decreasing with increasing energy,
as can be seen in Fig.~\ref{fig:MonoSimulations}. The emission angles were
found to be reproduced with an accuracy of about 1 degree.
Results obtained in this part were used to construct a response
matrix representing the distortions of the real spectrum induced
by the data analysis.

\begin{figure}
 \includegraphics[width = 0.8\columnwidth]{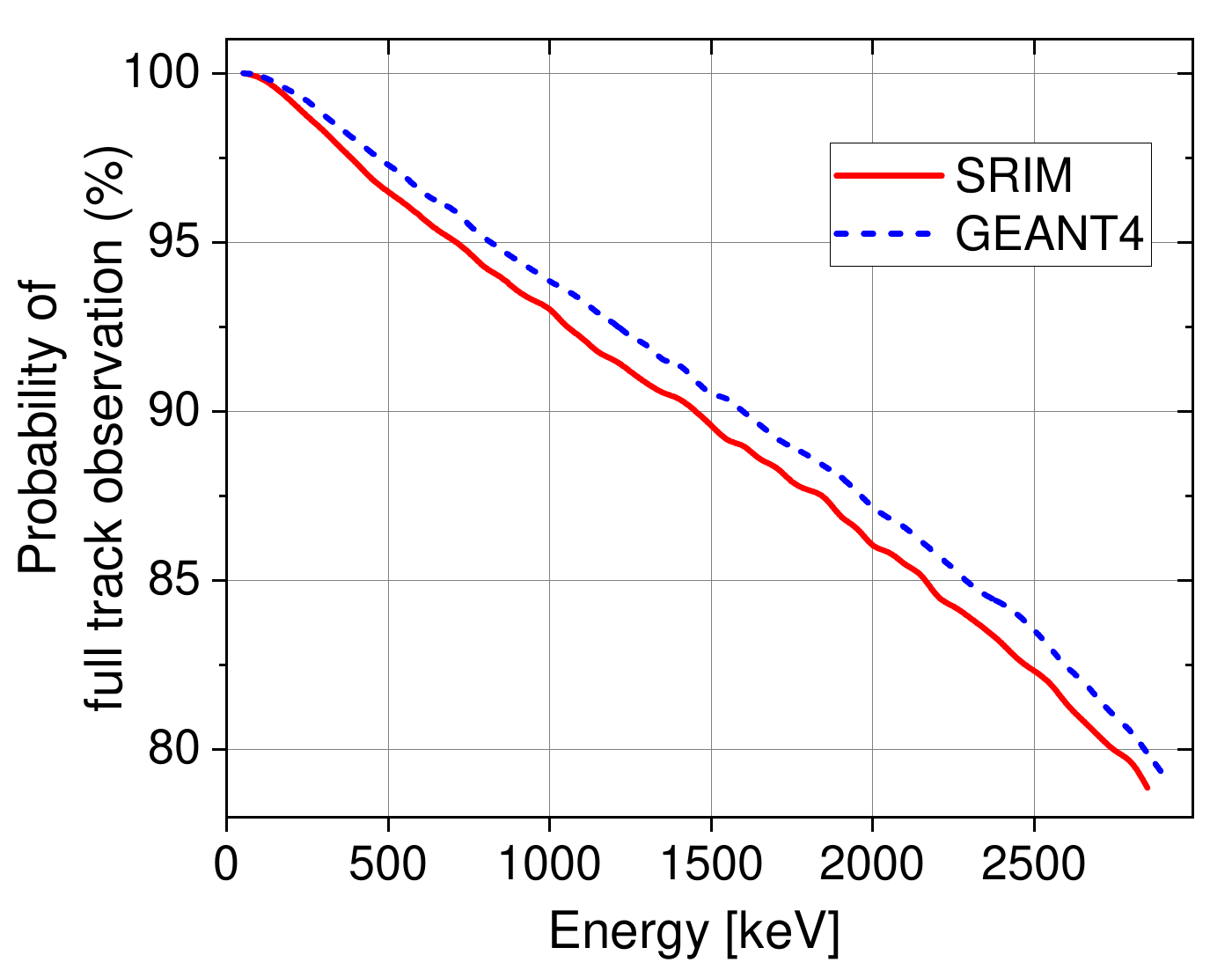}
 \caption{(Color online) Observation probability of the full track of a
 $\beta \alpha$ decay event in the OTPC detector in the HIE-ISOLDE experiment
 as a function of decay energy.}
 \label{fig:Efficiency}
\end{figure}

In the process of data reduction, as described above, decay events close
to the walls, as well as events found to run across one of the three gaps
in the active GEM area (see Fig.~\ref{fig:DecayPositionIsolde}), were
discarded. The probability of being affected by this process, however,
depends on the length of the track, and thus on the decay energy.
To figure out this dependence, other Monte Carlo simulations were
done. The distribution of decaying $^{11}$Be ions was assumed to
be composed of two clouds approximated by three-dimensional
Gaussian distributions. The width parameters of these distributions in the $x$
and $y$ directions were read out from the data (Fig.~\ref{fig:DecayPositionIsolde}).
The width in the $z$ direction was assumed to be the same as in the
$x$ direction. For a given decay energy, the decay location was sampled
randomly from this distribution, together with the emission angle
assuming isotropic decay. Then, taking into account the detector
dimensions, it was checked whether the event would be rejected because
of proximity to a wall or because of coming into a dead gap in the
GEM foil. This procedure was executed separately for the SRIM and the
GEANT4 energy loss models. The probability to observe the full event,
determined in such a way, as a function of decay energy, is shown
in Fig.~\ref{fig:Efficiency}.

\section{Results}


\subsection{Spectrum of $\beta$-delayed $\alpha$ particles}

\begin{figure}
 \includegraphics[width = \columnwidth]{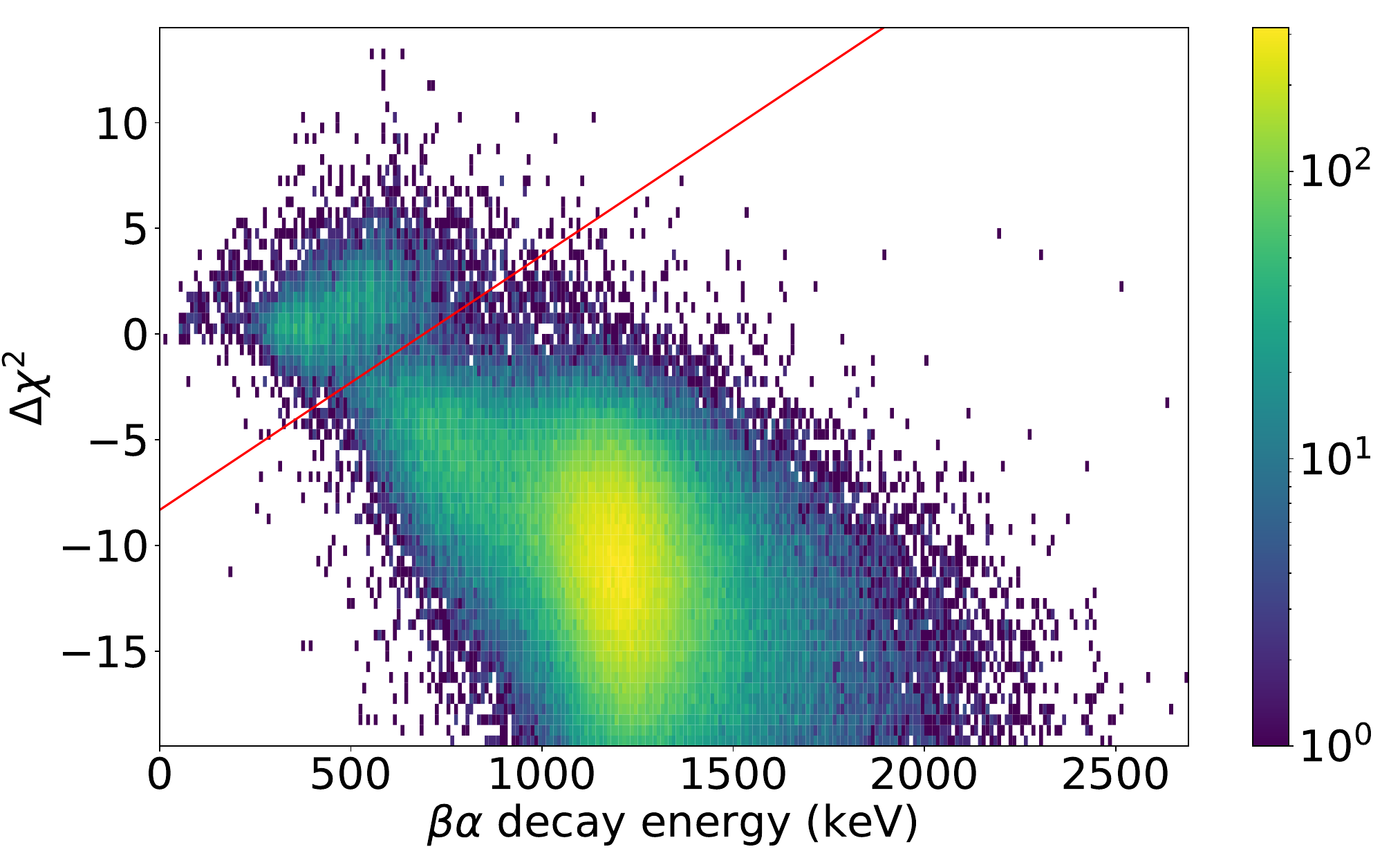}
 \caption{(Color online) The difference between the minimised chi-square values
 obtained in the reconstruction procedure of each event assuming the $\beta \alpha$
 and the $\beta p$ decay ($\Delta \chi ^2 = \chi_{\alpha}^{2} - \chi_{p}^{2}$),
 as a function of decay energy for the $\beta \alpha$ scenario.
 The SRIM energy-loss model was used. Only events located below the red line were
 included in the $\beta \alpha$ spectrum.}
 \label{fig:Delta_Chi2}
\end{figure}

Each event was reconstructed using both $\beta \alpha$ and $\beta p$
scenario yielding the minimized chi-square values, $\chi_{\alpha}^{2}$ and
$\chi_{p}^{2}$, respectively. In Fig.~\ref{fig:Delta_Chi2} the chi-square
difference, $\Delta \chi ^2 = \chi_{\alpha}^{2} - \chi_{p}^{2}$, as a function
of decay energy for the $\beta \alpha$ decay and for the SRIM model is shown.
Events having $\Delta \chi ^2 < 0$ were fitted better as representing $\beta \alpha$
process, while those with $\Delta \chi ^2 > 0$ are better reconstructed as $\beta p$
events. In Fig.~\ref{fig:Delta_Chi2} we can see a well separated group of events
at low energy with values $\Delta \chi ^2 \gtrsim 0$. It contains about 7 thousand
events, which corresponds to about 3\% of all statistics. Thus, they cannot represent
entirely the delayed proton emission because of their number and also because of
the energy, exceeding the $Q_p$ value.
All these events were inspected one-by-one and it was found that most
of them were showing a clear sign of damage, most often a cut. We believe that they
represent $\beta \alpha$ decays which occurred close to the cathode or to the GEM section.
Some of them could originate from $^{11}$Be ions that
were not fully neutralized and drifted to the cathode before decaying.
Many of these events were cut from both sides, suggesting a decay between the GEM foils.
The reason for such events being fitted better as $\beta p$ decays is that when the
energy deposition profile of the $\beta \alpha$ event is cut, it may become
similar to the proton emission profile (see Fig.~\ref{fig:ModelEvents}).
The red line in Fig.~\ref{fig:Delta_Chi2} marks the bottom of the ``valley'' between
two groups of events given by the local minimum of counts.
We decided to discard all events located above the red line in Fig.~\ref{fig:Delta_Chi2}
from the analysis of the $\beta \alpha$ spectrum.

\begin{figure}
 \includegraphics[width = 0.8\columnwidth]{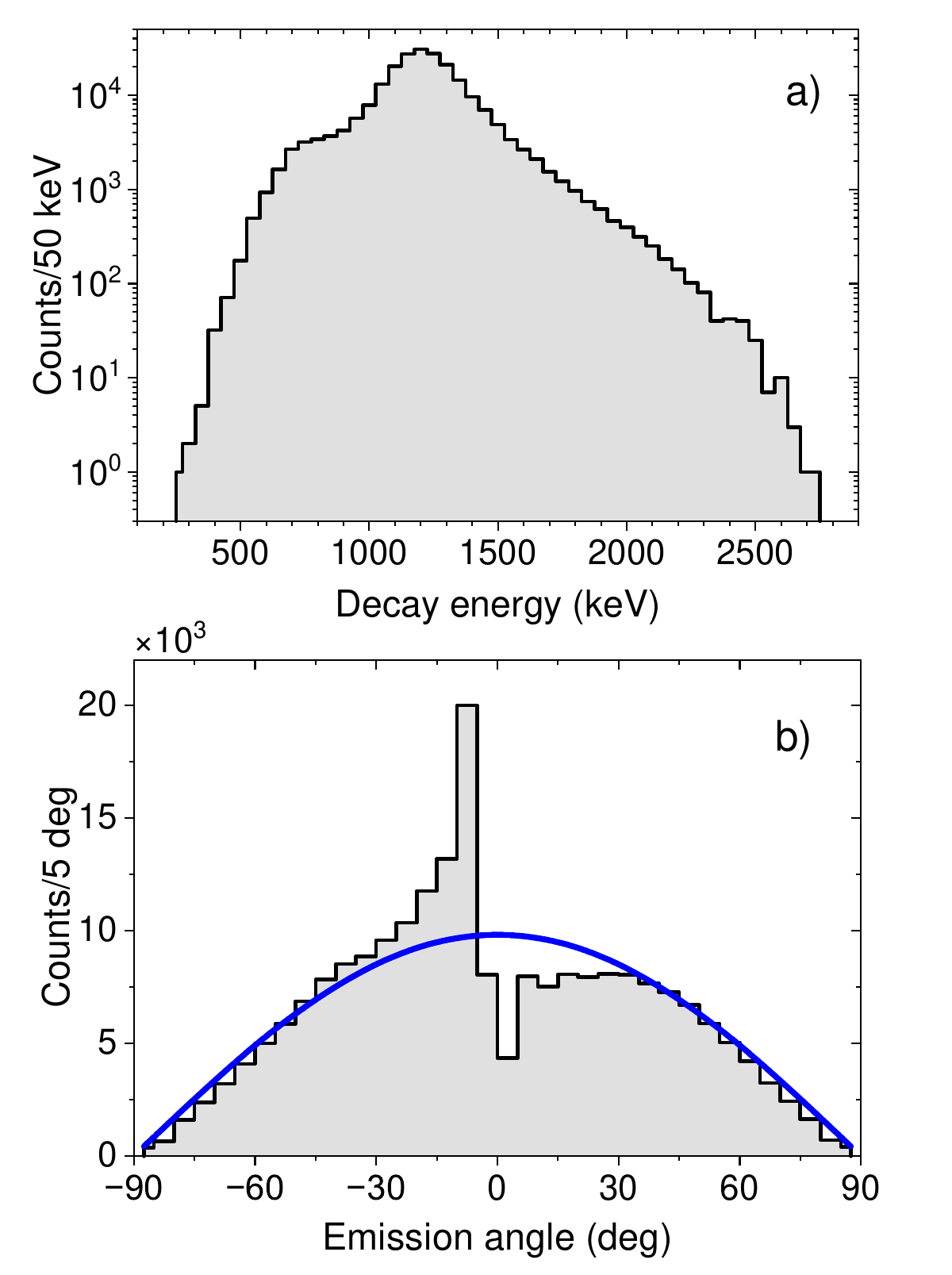}
 \caption{Reconstruction results of $\beta \alpha$ events for the SRIM
 energy-loss model. a) Energy spectrum, b) corresponding distribution of the emission angle.
  The smooth line shows the cosine function, expected for isotropic emission.  }
 \label{fig:ExpResults}
\end{figure}

The distributions of reconstructed decay energy and emission angle for
accepted 225 482 $\beta \alpha$ events, using the SRIM model, are shown in Fig.~\ref{fig:ExpResults}.
As long as decays occur within the detector volume, emission of delayed particles
should be isotropic. Indeed, the measured angular distribution is very close to
isotropic for the absolute value of the emission angle above $30^{\circ}$,
see Fig.~\ref{fig:ExpResults}b. A problem appears, however, at small angles:
there are too many events at small negative angles, while there are clearly
missing events with small positive angles. When particles are emitted at
a small angle i.e. almost parallel to the $x,y$ plane, the information on the angle
is encoded in the details of the PMT signal. For the zero angle this signal
should be symmetric and Gaussian-like. Then, however, the PMT sees a strong,
short pulse of light and nonlinearities in the charge processing slightly distort
the shape of the output signal. This is interpreted by the reconstruction
procedure as resulting from a small, non-zero angle. For larger angles,
the shape details are much less important, as the lengths of the
horizontal and vertical components carry the main information
about the angle. The erroneous determination of small angles should have a small
influence on the event energy, as in such case it is encoded mainly in the length
of the track which to the first order does not depend on the angle.
Nevertheless, in the analysis of the energy spectrum, we do check how
the final results are affected by the removal of events with incorrect
angles.
Note that with GEANT4 isotropically simulated events, where no PMT signal
distortions are taken into account, the angular distribution reconstructed with
the same procedure as experimental data did not present any distortion
at small angles.

In addition, the whole reconstruction procedure of accepted $\beta \alpha$ events
was repeated using the GEANT4 energy-loss model. The distribution of $\chi^2_{\alpha}$
was found broader than in the case of the SRIM model, indicating worse
reconstruction quality in general. The resulting energy spectrum
was found to have a very similar shape to the one stemming from the SRIM
model, although shifted by about 15 keV towards lower energies.
The angular distribution was found also asymmetric, very much like the one shown in
Fig.~\ref{fig:ExpResults}b, however, reversed with respect to zero angle.
Due to small differences in the range curves (Fig.~\ref{fig:Ranges}), the
predicted energy deposition profiles by the GEANT4 model are slightly different
than those from SRIM. As a result, the asymmetries present in the PMT
signal for events with an angle close to zero are interpreted by the
GEANT4 model in the opposite way. In the following, we will analyse only
the $\beta \alpha$ spectrum obtained with the SRIM model.

\subsection{$R$-matrix analysis}

The $\beta \alpha$ spectrum was analysed within the $R$-matrix framework \cite{Lane:1958},
following the approach taken by Refsgaard et al. \cite{Refsgaard:2019}.
We consider only Gamow-Teller transitions to $1/2^+$ and $3/2^+$ levels in $^{11}$B,
denoted in the following by the index $\lambda$. Each level is characterised by its
energy $E_{\lambda}$, $\beta$-decay feeding factor $B_{\lambda}$, and its reduced
width amplitudes $\gamma_{\lambda c}$, which are considered as free parameters in the
fitting procedure. The decay channel, $c$, denotes the final state in $^{7}$Li
($c=1$ for the ground state and $c=2$ for the $1/2^-$ excited state at 478 keV).
The decay energy spectrum is given by \cite{Barker:1988}:
\begin{eqnarray} \label{eq:widmo}
\nonumber  N(E)   & = & \sum_{c}N_c(E), \\
       N_c(E) & = & f_{\beta} \, P_c \, \Big| \sum_{\lambda \mu} B_{\lambda}\gamma_{\lambda c}A_{\lambda \mu} \Big|^2,
\end{eqnarray}
where
$f_{\beta}$ is the phase space factor for the $\beta$ decay, $P_c$ is the barrier
penetrability \cite{Lane:1958}, and $A_{\lambda \mu}$ is the level matrix defined in Ref. \cite{Brune:2002}.
The phase-space function $f_{\beta}$ was calculated with the LOGFT tool provided by the
NNDC portal \cite{NNDC:LOGFT}. For the Coulomb wave functions, necessary to calculate the
$P_c$ and $A_{\lambda \mu}$ terms, we used the formulation given in Ref.~\cite{Boersma:1969}.
For the channel radius parameter, we adopted $r_0 = 1.6$~fm and we took into account
only $p$-wave $\alpha$ emission ($l=1$) following Ref~\cite{Refsgaard:2019}.

For given values of fitting parameters, the model spectrum was calculated with Eq.~\ref{eq:widmo}
and corrected by the analysis response matrix and for the probability of observing the
full track (Fig.~\ref{fig:Efficiency}). Then, the Poisson likelihood chi-square value
was computed \cite{Baker:1984}:
\begin{equation} \label{eq:chi-square}
\chi^2_L = 2\sum_{i} \left[y_i - n_i + n_i\log\left(\frac{n_i}{y_i}\right)\right],
\end{equation}
where
$n_i$ and $y_i$ are the number of counts in the $i$-th data bin and the number of counts
predicted by the model for this bin, respectively, and the sum runs over all data bins.
The minimization of $\chi^2_L$ was done using MINUIT2 routines provided by the
Phython \verb"iminuit" package \cite{minuit}.

\begin{table}
\begin{center}
\caption{Four major $R$-matrix models fitted to the $\beta \alpha$ spectrum of $^{11}$Be.
Minimized value of $\chi^2_L$ divided by the number of degrees of freedom is shown in
the second column.   }
\vspace*{0.2cm}
\begingroup
\renewcommand{\arraystretch}{1.2}
\begin{tabular}{ c | c }
  \hline  \hline
  Model & $\chi^2_L$/ndf  \\
  \hline \hline
  $3/2^+$ + $1/2^+$ & 13.81   \\
  $3/2^+$ + $3/2^+$ & 5.03   \\
  $3/2^+$ + $3/2^+$ + $1/2^+$ & 2.21  \\
  $3/2^+$ + $3/2^+$ + $3/2^+$ & 3.10 \\
  \hline \hline
\end{tabular}
\endgroup
\label{tab:MainModels}
\end{center}
\end{table}

First, using the full spectrum obtained with the SRIM energy-loss model, we
considered a few variants of the $R$-matrix model differing by the assumed
levels through which the $\beta$ decay proceeds. They are listed in Table~\ref{tab:MainModels}
together with the minimized value of $\chi^2_L$ per number of degrees of freedom.
We reproduce the observation made in Ref.~\cite{Refsgaard:2019} that while
two levels are concerned, the model with two $3/2^+$ states reproduces the
measured spectrum significantly better than the model with $3/2^+$ and $1/2^+$ states.
Since in the level scheme of $^{11}$B (Fig.~\ref{fig:DecayScheme}) there are three
levels that could be involved in the $\beta \alpha$ process, we tested also
three-level scenarios. Adding the third $3/2^+$ level does improve the quality of
the fit. However, if we assume the third level to be $1/2^+$, as tentatively assigned
to the state at 9820~keV \cite{Kelley:2012}, we obtain the best overall description of the data.
Thus, for further consideration, we take only the models with two $3/2^+$ levels
and two $3/2^+$ plus one $1/2^+$ levels.

\begin{table}[hb]
\begin{center}
\caption{Results of different variants of the two selected
$R$-matrix models fitted to the $\beta \alpha$ spectrum of $^{11}$Be.
The second column indicates 
if the full experimental spectrum was taken or the one with removed events having
small emission angles, see text for details. Level energies are in keV. Numbers
in parentheses denote statistical errors.}
\vspace*{0.2cm}
\begingroup
\renewcommand{\arraystretch}{1.2}
\begin{tabular}{ c | c | c | c | c | c }
  \hline \hline
  Model & Variant & $\chi^2_L$/ndf & $E_1$ & $E_2$& $E_3$ \\
  \hline \hline
  \multirow{2}*{$2 \times 3/2^+$} & full & 5.03 & 9906(1) & 11795(100) & - \\
   & removed & 3.02 & 9901(1) & 11682(75)& - \\
  \hline
  \multirow{2}{1cm}{$2 \times 3/2^+$ \\ + $ 1/2^+$}& full & 2.21 & 9923(4) & 11817(100) & 9813(20)\\
   & removed & 1.64 & 9912(6) & 11672(200) & 9810(25) \\
  \hline \hline
\end{tabular}
\endgroup
\label{tab:SelectedModels}
\end{center}
\end{table}

\begin{figure}
 \includegraphics[width = \columnwidth]{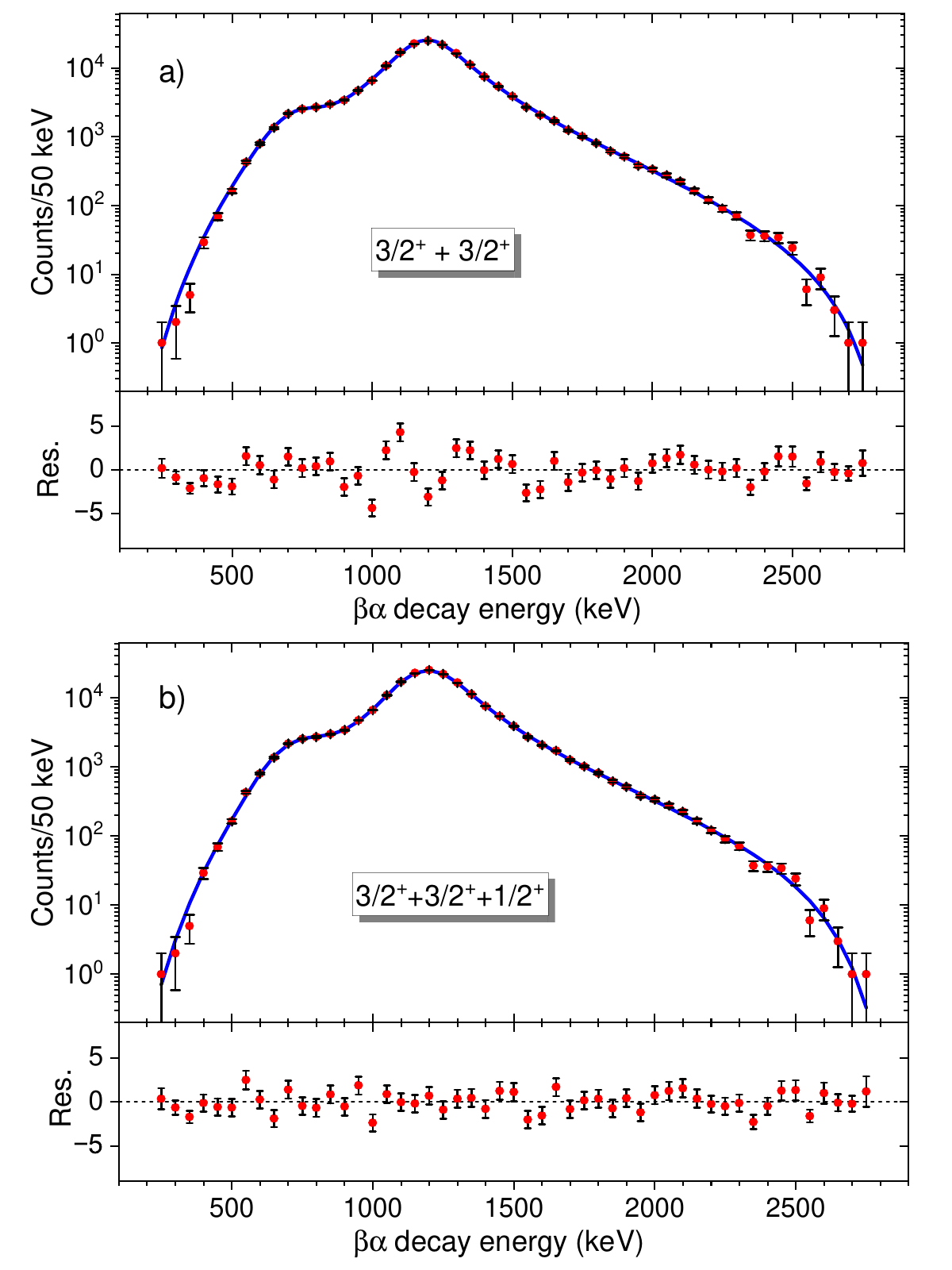}
 \caption{(Color online) Measured $\beta \alpha$ spectrum of $^{11}$Be
 (red points with error bars) compared to best $R$-matrix fits (blue line).
  Data come from the SRIM-model reconstruction with removed events having
  the emission angle in the range ($-20^{\circ}$, $-5^{\circ}$). In the bottom
  parts, the fit residuals, $(n_i - y_i)/\sqrt{y_i}$, are shown. The models assuming
  two $3/2^+$ (a) and two $3/2^+$ plus one $1/2^+$ (b) levels in $^{11}$B are presented.}
 \label{fig:BestFits}
\end{figure}

In the next step, we checked how the final results will change if we remove
from the SRIM experimental spectrum events having the emission angle in the range
($-20^{\circ}$, $-5^{\circ}$), which contains a large part of misidentified
angles, as shown in Fig.~\ref{fig:ExpResults}b. The results
of this exercise are shown in Table~\ref{tab:SelectedModels}.
It can be seen that for both R-matrix models the fit is improved
when the events having small negative emission
angles are removed, leaving 180 556 events in the spectrum.
Again, the best result overall is achieved for the
$R$-matrix model with two $3/2^+$ plus $1/2^+$ levels. The two best fits are shown
in Fig.~\ref{fig:BestFits}.


From the best-fitted parameters, in addition to the level energy, we can
deduce the level widths and quantities characterising $\beta$ decay of $^{11}$Be.
Instead of reduced width amplitudes $\gamma_{\lambda c}$, we present their ratio
to the Wigner limit \cite{Teichmann:1952}:
\begin{equation}
\theta^2_{\lambda c} = \frac{\gamma_{\lambda c}^2 \, \mu a_c^2 }{\hbar^2},
\end{equation}
where $\mu$ is the reduced mass of $^{7}$Li and $\alpha$ particle and $a_c$ is
the channel radius, $a_c = r_0 \, (A_1^{1/3}+A_2^{1/3})$. The observed width
of the level is given by \cite{Brune:2002}:
\begin{eqnarray}
\nonumber \Gamma_{\lambda} & = & \sum_{c} \Gamma_{\lambda c}, \\
\Gamma_{\lambda c} & = & \frac{2P_c\gamma^2_{\lambda c}}{1 + \sum_{c} \gamma^2_{\lambda c}\frac{dS_c}{dE}\Big|_{E_{\lambda}}},
\end{eqnarray}
where $S_c$ is the $R$-matrix shift function \cite{Lane:1958}.
The Gamow-Teller matrix elements can be determined approximately \cite{Barker:1988} from:
\begin{equation}
M_{GT,\lambda} = \left(\frac{\pi D}{Nt_{1/2}}\right)^{\frac{1}{2}} \left(1 + \sum_{c}\gamma^2_{\lambda c} \frac{dS_c}{dE}\Big|_{E_{\lambda}} \right)^{-\frac{1}{2}}B_{\lambda},
\end{equation}
where
$D = 6147(2)$~s \cite{Hardy:2005}, $t_{1/2}$ is the partial half-life for the $\beta \alpha$
decay, and $N$ is the number of counts in the spectrum. More often used quantities
$B_{GT}$ and $\log (ft)$ follow:
\begin{equation}
B_{GT, \lambda} = \left(\frac{g_A}{g_V}\right)^{-2}M_{GT,\lambda}^2; \,\,\,
\log(ft)_{\lambda}= \log \left[\frac{D}{M_{GT,\lambda}^2} \right],
\end{equation}
where $|\frac{g_A}{g_V}| = 1.2723(23)$ \cite{Tanabashi:2018}.
All results for the two best fits are listed in Table~\ref{tab:Results} together
with values published in Ref.~\cite{Refsgaard:2019} which corresponds to the
$R$-matrix model with two $3/2^+$ levels.

\begin{table}[h]
\begin{center}
\caption{Best fit parameters and derived quantities for two $R$-matrix models found
to best reproduce the $\beta \alpha$ spectrum determined in this work, compared
to the results of Ref.~\cite{Refsgaard:2019}. In parentheses, the statistical errors
are given, while in square brackets the systematical errors are shown. }
\vspace*{0.2cm}
\begingroup
\renewcommand{\arraystretch}{1.2}
\begin{tabular}{ c | c | c | c }
  \hline \hline
      & $2 \times 3/2^+$ Ref.~\cite{Refsgaard:2019} & $2 \times 3/2^+$ & $2 \times 3/2^+ + 1/2^+$ \\
  \hline \hline
  $E_1$ (keV) & 9 846(1)[10] & 9 901(1)[30] & 9 912(6)[35] \\
  $B_1/\sqrt{N}$ & 0.161(2) & 0.152(1)[2] & 0.140(10)[3]\\
  $\theta^2_{11}$  & 1.31(2) & 1.04(1)[17] & 0.92(6)[14]\\
  $\theta^2_{12}$ & 0.84(2) & 0.44(1)[13] & 0.42(3)[14]\\
  $\Gamma_{11}$ (keV) & 233(3)[3] & 263(2)[4]  & 251(4)[7] \\
  $\Gamma_{12}$ (keV)& 20.4(3)[3] & 18.9(3)[2] & 20(1)[1] \\
  $M_{GT_1}$ & 0.717(12)[7] & 0.760(2)[40] & 0.714(20)[25]\\
  $B_{GT_1}$ & 0.318(11)[6] & 0.357(2)[35] & 0.315(15)[20]\\
  $\log(ft)_{1}$ & 4.08(3)[2] & 4.027(2)[40] & 4.08(2)[3]\\
  \hline
  $E_2$ (keV) & 11 490(80)[50] & 11 682(75)[260]  & 11 672(200)[40]\\
  $B_2/\sqrt{N}$ & 0.156(26) & 0.160(4)[70] & 0.09(4)[20]\\
  $\theta^2_{21} \footnotemark[1] $ & -0.21(7)  & -0.152(25)[60] & -0.39(13)[30]\\
  $\theta^2_{22} \footnotemark[1] $ & 0.029(37) & 0.015(16)[25] & -0.01(5)[5]\\
  $\Gamma_{21}$ (keV) & 430(150)[50] & 338(64)[120] & 854(200)[670]\\
  $\Gamma_{22}$ (keV) & 50(60)[50] & 27(28)[30] & 18(50)[90]\\
  $M_{GT_2}$ &  1.05(17)[5] & 1.08(3)[50] & 0.63(13)[120]\\
  $B_{GT_2}$ & 0.7(2)[1] & 0.72(4)[80] & 0.25(10)[200]\\
  $\log(ft)_{2}$ & 3.8(3)[1] & 3.72(2)[30] & 4.2(2)[10]\\
  \hline
  $E_3$ (keV) & & & 9 810(25)[40]\\
  $B_3/\sqrt{N}$ & & & 0.042(22)[15] \\
  $\theta^2_{31}$  &    &   & 0.61(27)[10]\\
  $\theta^2_{32}$ &   &   & 0.33(3)[15]\\
  $\Gamma_{31}$ (keV) & & & 146(32)[25] \\
  $\Gamma_{32}$ (keV) & & & 9(3)[6]\\
  $M_{GT_3}$ & & & 0.23(5)[6]\\
  $B_{GT_3}$ & & & 0.032(15)[20] \\
  $\log(ft)_{3}$ & & & 5.1(2)[2] \\
  \hline \hline
\end{tabular}
\footnotetext[1]{The sign in these entries indicates the sign on the corresponding
reduced width amplitude, $\gamma_{\lambda c}$.}
\endgroup
\label{tab:Results}
\end{center}
\end{table}

Statistical errors given in Table~\ref{tab:Results} were calculated using
covariance matrices provided by the fitting package. The main systematic error
in our analysis is related to energy determination. It has two sources.
One comes directly from the energy-loss models. Some hint of this contribution
is given by the difference between the SRIM and the GEANT4 models, which
amounts to $\sim15$~keV at the decay energy of 1200~keV. The second source is the
uncertainty of the OTPC gas density which is affected by changes of the
temperature and pressure, and the inaccuracy of the gas-flow rate from which
the gas composition is determined. The estimated  overall accuracy of the
gas density is about 1\% and this leads to an energy shift of 30~keV
for a $\beta \alpha$ event at 1200~keV. We adopted this value as the
measure of systematic error of energy and we assumed that it grows linearly
with increasing energy. Then we changed the bin energies
correspondingly and refitted the spectra with the $R$-matrix models. The resulting
changes of the fit parameters were used to determine the systematic
errors shown in Table~\ref{tab:Results} in square brackets.

Our results support one of the main conclusions of Ref.~\cite{Refsgaard:2019},
namely that the satisfactory description of the $\beta \alpha$ spectrum from $^{11}$Be
requires at least two $3/2^+$ states. Moreover, the $\beta$-decay strengths,
expressed by the $B_{GT}$ or $\log (ft)$ values, coincide within the error
bars with the values determined in Ref.~\cite{Refsgaard:2019}.
The energy of $E_1$, the state dominating the $\beta \alpha$ emission,
and its width are very close to the values given in \cite{Refsgaard:2019},
although they are a bit larger. Our value for $E_1$ is consistent with the
known $3/2^+$ level in $^{11}$B at 9873~keV (see Fig.~\ref{fig:DecayScheme}).
The width of this state, however, appears almost three times
larger than 109(14)~keV measured for the 9873~keV state in scattering and
reaction experiments \cite{Kelley:2012}. The energy of our second
$3/2^+$ level suggests that it could be the known state at 11450~keV
tentatively assigned as $3/2^+$. Unfortunately, our value has a large
systematical uncertainty resulting from the rather small contribution
to the spectrum. The measured width for this state is
smaller than the one found in Ref.~\cite{Refsgaard:2019} but it is still
more than three times larger than 93(17)~keV adopted for the 11450~keV
level \cite{Kelley:2012}.
Whether the two states we see in the $\beta \alpha$ emission do coincide
with the 9873~keV and 11450~keV levels requires further, independent
investigations.

The quality of the $R$-matrix fit improves if we add a $1/2^+$ state.
As a result of this step, all parameters of the first state and the
energy of the second state do not change significantly.
The width of the second state increases a lot and its $\beta$ feeding
decreases, but they are not well determined having large errors.
The third state appears with a width of the order
of 150~keV and its energy of 9810~keV is close to the known
state in $^{11}$B at 9820(25)~keV which is tentatively assigned
as $1/2^+$ (see Fig.~\ref{fig:DecayScheme}). Up to now, the only evidence
for this state came from quasi-elastic electron scattering on
$^{12}$C \cite{Steenhoven:1988}. Independent confirmation of the
energy, spin, and width of this state would be helpful. It is important
to note that, as can be seen  in Fig.~\ref{fig:BestFits}, the improvement
of the $R$-matrix fit by adding the $1/2^+$ state comes mainly from
the energy region 900--1400~keV. The spectrum in this region is not
affected by our, somewhat arbitrary, selection of good $\beta \alpha$ events.

Having the best-fitted $R$-matrix models, we can determine the branching
ratio for the two $\beta \alpha$ decay channels by integrating their
contributions to the final spectrum over the whole energy range.
For the transition to the $^{7}$Li ground state and to the excited
state at 478~keV, we found the branching ratios to be $93.7(4)$\%
and $6.3(4)$\%, respectively, in the two $3/2^+$ levels scenario.
For the model with the three levels ($3/2^+ + 3/2^+ + 1/2^+$), the corresponding values
are $92.9(3)$\% and $7.1(3)$\%. These numbers are close to the
results of Ref.~\cite{Refsgaard:2019} which are $92.1(3)$\%
and $7.9(3)$\%, respectively.

\subsection{Search for $\beta p$ decay channel}

Candidates for the $\beta$-delayed proton emission were looked for among
those events which were better fitted as $\beta p$ events than $\beta \alpha$
in the reconstruction procedure. More precisely, all events that were
discarded from the $\beta \alpha$ spectrum, i.e. those above the red
line in Fig.~\ref{fig:Delta_Chi2}, were taken into account.
They consist predominantly of events with $\Delta \chi ^2 > 0$  but they
include also events with $\Delta \chi ^2 \lesssim 0$. The latter have to be
considered because at low energy the track profiles of the two types of particle
decays ($^{7}$Li$ + \alpha$ and $^{10}$Be$ + p$ ) become similar to each other
and it is hard to tell them apart. As reported above, all these events
were carefully inspected one by one and those which bore clear signs
of damage were removed. Among the removed events, the majority represented
a track apparently cut, thus indicating a particle hitting the cathode or the
GEM section and thus depositing only a part of its energy in the detector.
Another numerous category contained events of equal, sharply defined vertical
length, thus suggesting a decay between the GEM foils. In addition, we observed
some events with other types of distortions, like a double-bump structure,
or resulting apparently from a spark in the GEM structure.
All removed events looked very different than expected for the $\beta$-delayed
emission of protons or $\alpha$ particles.
We observed also a number of imperfect events, for which it was not obvious whether
they should be retained or removed. These events were used to estimate the
uncertainty of the selection procedure.

\begin{figure}[h]
 \includegraphics[width = 0.9\columnwidth]{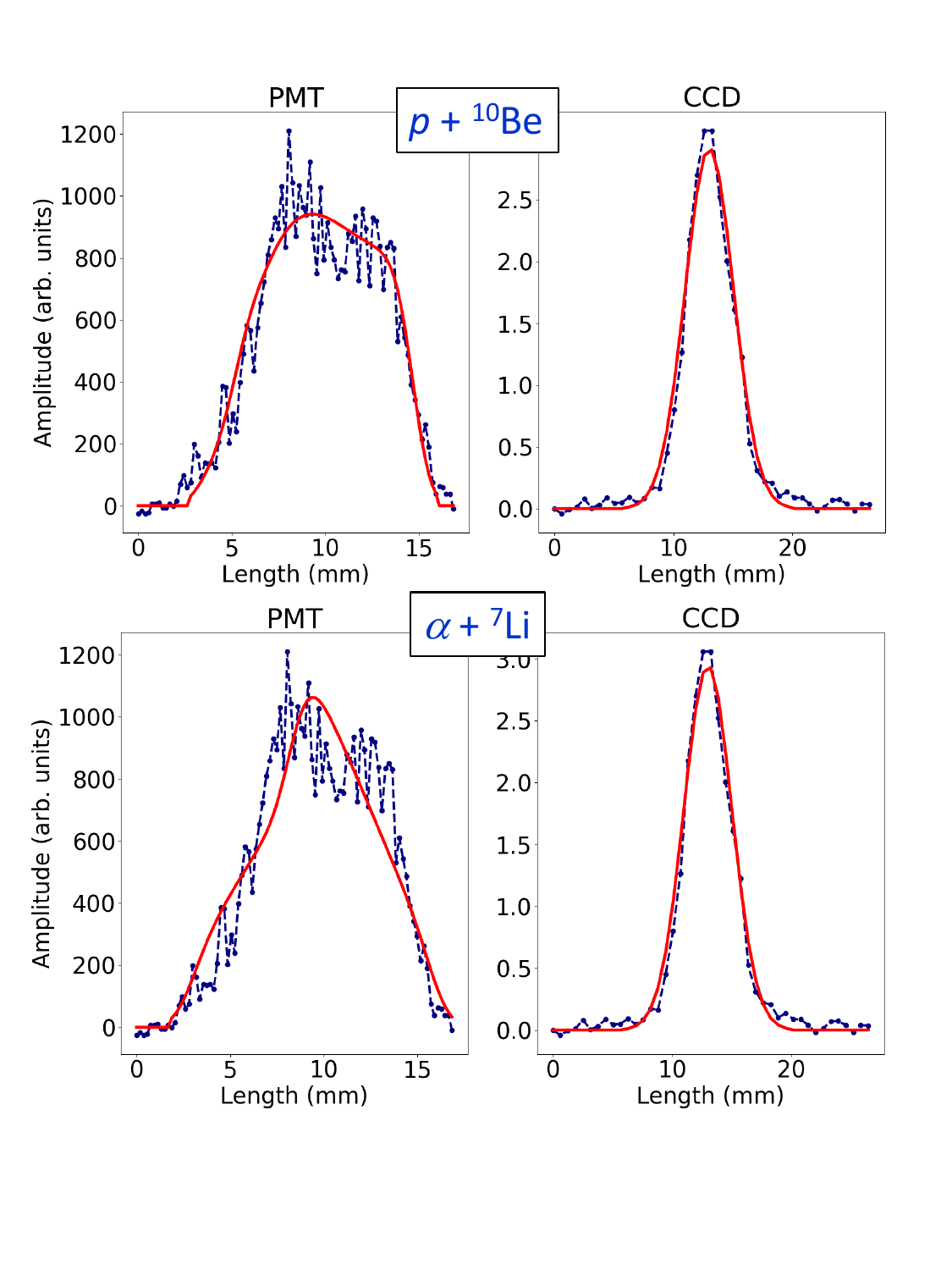}
 \caption{(Color online) Example event from the group that can contain the $\beta p$ events.
  The blue points represent the data, the red lines show the best fit of the model signal.
  In the top row, the model corresponds to a $\beta p$ emission with the decay energy of 195~keV
  and the emission angle of the proton of -83$^{\circ}$, while in the bottom row the best-fitted $\beta \alpha$
  model is shown with the decay energy of 229~keV and the emission angle of the $\alpha$ particle of 87$^{\circ}$.
}
 \label{fig:ProtonExample}
\end{figure}

\begin{figure}[h]
 \includegraphics[width = 0.9\columnwidth]{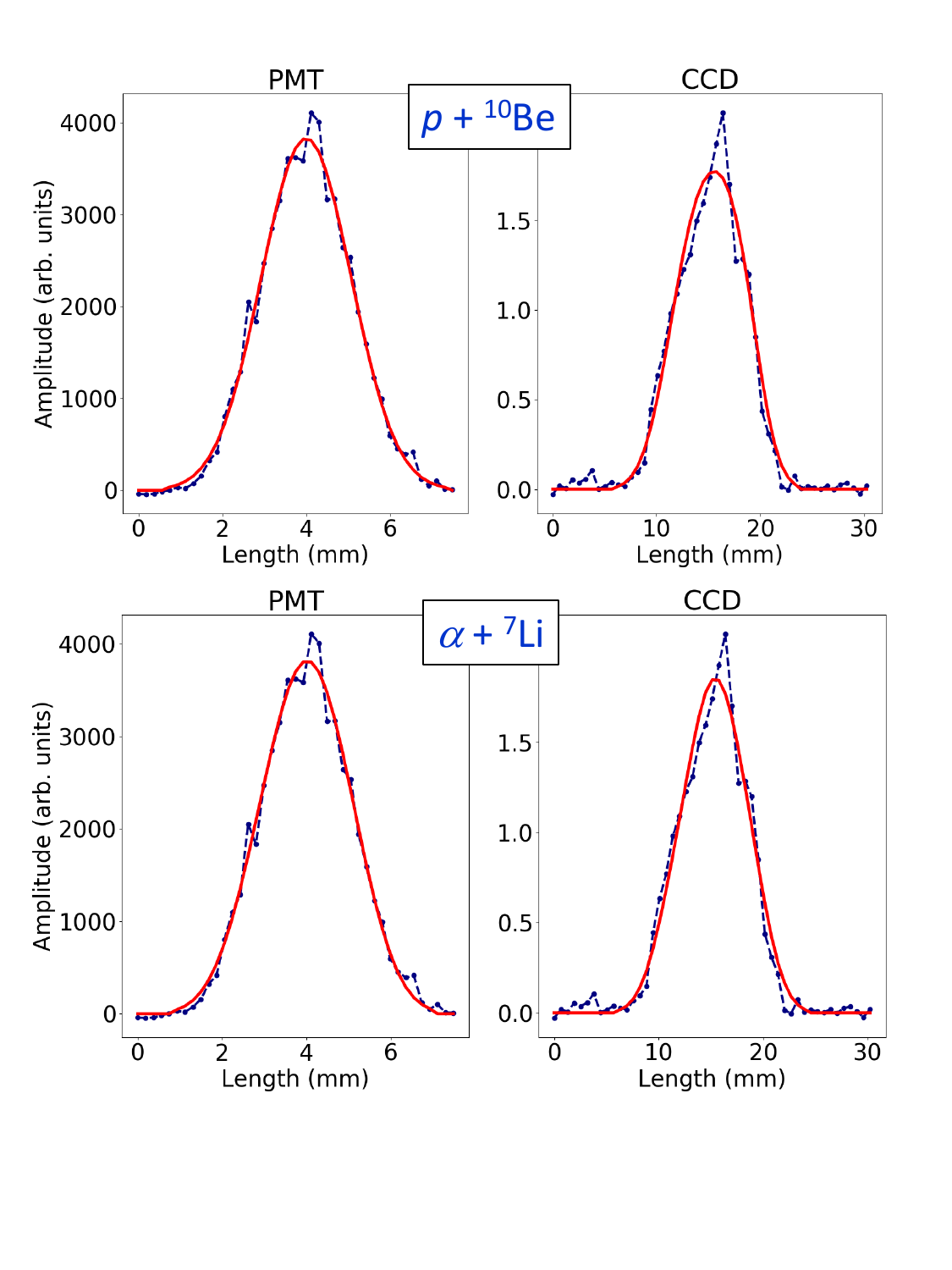}
 \caption{(Color online) The same as Fig.~\ref{fig:ProtonExample} but for a different event.
  In the top row, the model corresponds to a $\beta p$ emission with the decay energy of 176~keV
  and the emission angle of the proton of 8$^{\circ}$, while in the bottom row the best-fitted $\beta \alpha$
  model is shown with the decay energy of 190~keV and the emission angle of the $\alpha$ particle of 17$^{\circ}$.
}
 \label{fig:ProtonAlphaExample}
\end{figure}

After this selection, about 2200 events remained as candidates for
$\beta p$ events. Nevertheless, they also can represent $\beta \alpha$
events that appear very similar to the $\beta p$ ones and/or for which
the effect of damage was smeared out by the electron diffusion.
For illustration, we present two examples from this group.
In the first, shown in Fig.~\ref{fig:ProtonExample}, the model of the $\beta p$
emission fits the data clearly better than the $\beta \alpha$ decay scenario.
In contrast, the event shown in Fig.~\ref{fig:ProtonAlphaExample} is fitted
equally well by the two alternative decay modes.



\begin{figure}
 \includegraphics[width = 1.0 \columnwidth]{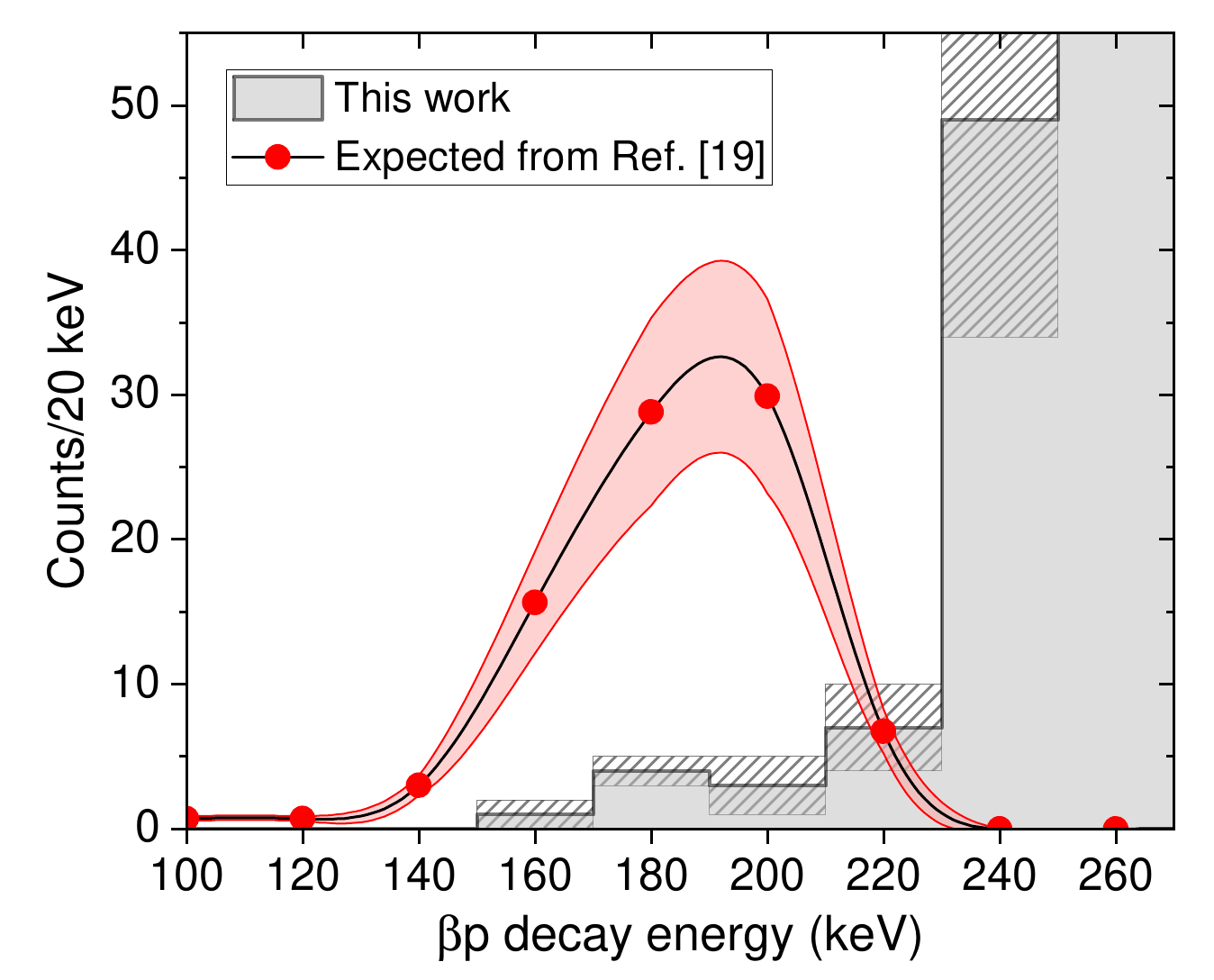}
 \caption{(Color online) Decay energy spectrum for $\beta p$ candidate events.
  The grey histogram shows the data from this work, with the line pattern indicating
  the uncertainty of the selection procedure. The red points represent the
  spectrum expected if the $\beta$-delayed proton emission in $^{11}$Be
  proceeded as reported in Ref.~\cite{Ayyad:2019}. The red band
  illustrates the uncertainty of the claimed  $\beta p$ branching
  ratio, see text for details. }
 \label{fig:Protons}
\end{figure}

The low-energy part of the energy spectrum of $\beta p$ candidates, selected
as described above, is shown in Fig.~\ref{fig:Protons} with a grey histogram
where the uncertainty of the selection procedure is marked with a line pattern.
Interestingly, the number of counts in the region around 200~keV is very small.
In the work of Ayyad et al.~\cite{Ayyad:2019} the $\beta$-delayed
protons emitted from $^{11}$Be were reported to be observed in a narrow
peak at a decay energy of 196~keV. This peak was interpreted as
originating from a narrow resonance in $^{11}$B having a width of 12(5)~keV.
The resulting branching ratio for the $\beta p$ emission was determined as
$(1.3 \pm 0.3) \times 10^{-5}$~\cite{Ayyad:2019}.

To compare our results with the observation of Ref.~\cite{Ayyad:2019},
we run the GEANT4 simulations in order to see how the reported $\beta p$
process would show up in our spectrum. A number of $\beta p$ events
were generated with the decay energy having a Gaussian distribution
centered at 196~keV with a variance of 12~keV. These events were
reconstructed in the same way as the real data were, using the GEANT4
energy-loss model for consistency.
The obtained spectrum was then normalized to the number of counts
in the full $\beta \alpha$ spectrum (shown in Fig.~\ref{fig:ExpResults}),
taking into account the branching ratios of $3.3 \times 10^{-2}$,
and $1.3 \times 10^{-5}$ for the $\beta \alpha$ and $\beta p$ emission,
respectively. The result is shown in Fig.~\ref{fig:Protons} with the
red points while the red band represents the 23\% uncertainty
of the $\beta p$ branching ratio determined in Ref.~\cite{Ayyad:2019}.
In total, we should see about ($90 \pm 20$)
$\beta p$ events which is clearly not the case.

For a decay energy below 230~keV, we see in total $15 \pm 4$ counts.
Some of them may belong to the tail of the $\beta \alpha$
background visible in Fig.~\ref{fig:Protons} and there is no
unambiguous way to distinguish a good $\beta p$ event from the
distorted $\beta \alpha$ one in this group of events.
For example, we see in this group cases that fit well as
delayed proton emission, but with the decay energy above 350~keV.
Such events cannot represent $\beta p$ decay for which the maximal
energy is 281~keV.
Therefore, we can only conclude that for the energy range below 230~keV,
the number of $\beta p$ events in our spectrum is less or equal
to $15 \pm 4_{\rm stat} \pm 4_{\rm sys}$.
This yields an upper limit for the branching ratio of $\beta$-delayed
proton emission in $^{11}$Be of
$b_{\beta p} \leq (2.2 \pm 0.6_{\rm stat} \pm 0.6_{\rm sys}) \times 10^{-6}$.
This value agrees with the limit obtained from the recent indirect
measurement of Riisager et al.~\cite{Riisager:2020}, but is in
strong conflict with the result of Ayyad et al.~\cite{Ayyad:2019}.
Since the energy window for the $\beta p$ emission is opened up to
281~keV, it is possible that a number of $\beta p$ events
are hidden in the $\beta \alpha$ background above 230~keV.
We cannot exclude such a possibility, but we note that this would
also contradict the observation made in Ref.~\cite{Ayyad:2019}.

Eight events in the energy range 150 -- 210~keV may well represent
$\beta p$ decay events. If we tentatively make such an assignment,
the branching ratio would be
$b_{\beta p} = (1.2 \pm 0.4_{\rm stat} \pm 0.4_{\rm sys} ) \times 10^{-6}$.

\section{Summary and conclusions}

We have studied $\beta$ decay of $^{11}$Be with delayed emission of charged
particles using the Warsaw TPC detector with optical readout.
The $\beta$-delayed $\alpha$ emission was investigated in detail.
In the experiment carried out at the INFN-LNS the branching ratio
for this process was measured by counting incoming ions of $^{11}$Be
and decay events with emission of particles. From about 2000 decay
events recorded, the value of $b_{\beta \alpha} = 3.27(46) \times 10^{-2}$
was determined. This value agrees with the most recent and most
accurate value to date, $3.30(10)\times10^{-2}$~\cite{Refsgaard:2019},
although it has a larger error bar, mainly due
to the large uncertainty of the stopping probability of ions in the detector.

In the second experiment, made at the HIE-ISOLDE facility, the energy spectrum
of $\beta \alpha$ decay events was measured in the full energy window available
for this process. More than 200 000 counts were recorded and analysed.
For the first time the low-energy
part of this spectrum, below 500~keV, was obtained. Its detailed shape
at lowest energies could have been affected by a background due to distorted
signals, most probably because of decays in proximity to the cathode plate
or the amplification zone. Nevertheless, the spectrum was found to agree
well with the results reported in Ref.~\cite{Refsgaard:2019}.
The analysis in the framework of the $R$-matrix formalism supported
the conclusion that at least two $3/2^+$ levels in $^{11}$B are needed
to satisfactorily reproduce the $\beta \alpha$ spectrum~\cite{Refsgaard:2019}.
The energies and $\log {ft}$ values for these states were found consistent
with the results of Ref.~\cite{Refsgaard:2019} within the rather substantial
systematical errors. One of these states, dominating by far the observed spectrum,
at about 9900~keV, probably corresponds to the known $3/2^+$ state
at 9873~keV, although the width assigned to it \cite{Kelley:2012}
is about 2.5 times narrower than the one observed by us and in
Ref.~\cite{Refsgaard:2019}. The energy of the second level, at about 11700~keV
was determined with a large systematical error. However, within this
uncertainty, it fits to the known state at 11450~keV, tentatively assigned
to be a $3/2^+$~\cite{Kelley:2012}. But again, the width deduced by us, as well as
in Ref.~\cite{Refsgaard:2019}, is more than 3 times larger than the one adopted for the
11450~keV state \cite{Kelley:2012}.

We found that a better description of the measured $\beta \alpha$ spectrum
in the $R$-matrix model can be achieved by taking into account a third state of
$1/2^+$ nature at about 9800~keV. It appears to be narrower ($\Gamma \approx 150$~keV)
than the other two levels. These findings fit to the known state at 9820~keV
tentatively assigned as $1/2^+$. Thus, the best description of the
measured $\beta \alpha$ spectrum with the $R$-matrix model is
achieved with the pattern of $3/2^+$ and $1/2^+$ states which is very
similar to the known level scheme of $^{11}$B \cite{Kelley:2012}.

The general agreement of our description of the $\beta \alpha$ spectrum
from $^{11}$Be with the findings of Ref.~\cite{Refsgaard:2019} gives us
confidence that our detection technique based on the OTPC detector
is well suited for spectroscopic studies of $\beta$-delayed charged
particle emission even in case of very large number of events. In contrast
to experiments using silicon detectors, where the energy is extracted from
the signal amplitude only, in our method, each event is reconstructed in
detail separately and the energy is determined both from the track length and
the energy-deposit distribution along the track. Both approaches have their
advantages and both suffer from different systematical errors, thus they should
be considered as complementary.

The main goal of this project was to search for the $\beta$-delayed proton
emission from $^{11}$Be. This task was hampered by the presence of a background
posed by distorted $\beta \alpha$ events. However, after careful inspection
of all events, only 15 of them were found in the energy range where
about 90 $\beta p$ events were expected according to Ref.~\cite{Ayyad:2019}.
From this number, we deduced the upper limit for the $\beta p$ branching ratio
to be $(2.2 \pm 0.6_{\rm stat} \pm 0.6_{\rm sys}) \times 10^{-6}$
for the energy below 230~keV.
This value is in conflict
with the results reported in Ref.~\cite{Riisager:2014} and in Ref.~\cite{Ayyad:2019}.
On the other hand, our result agrees with the final limit determined by Riisager
et al.~\cite{Riisager:2020}, though we note that the latter is based on
internally conflicting results.


The final answer to the question of whether in the decay of $^{11}$Be are
protons emitted, and if yes, with which probability, still needs further
investigation. A gaseous TPC-like detector seems to be an optimal tool
for such studies. For better separation of $\beta \alpha$ events
from $\beta p$ ones, a thinner gas mixture, thus at lower pressure
than the one used in the present study would be advantageous, because
for longer tracks the differences between the energy-deposition profiles
are more pronounced. To minimize the background from distorted $\beta \alpha$
events, bunches of $^{11}$Be ions should be implanted in the
center of the active volume and the time between consecutive bunches
should be large enough for all nuclei to decay in between. These measures
would reduce the statistics which could be counterbalanced by using a
TPC detector with an electronic readout, where the drift-time waveform
is recorded independently for different regions (pads) of the anode.
This feature would allow observation of events which are close in time,
provided they are sufficiently separated in space.

\section*{Acknowledgements}

We would like to thank the INFN-LNS staff and the ISOLDE Collaboration and
technical teams for their support during the experiments and for the excellent
quality of the $^{11}$Be beam.
This work was supported by the Polish Ministry of Science and Higher Education
under Contract No. 2021/WK/07, by the National Science Center, Poland,
under Contracts No. 2019/33/B/ST2/02908 and No. 2019/35/D/ST2/02081,
by the Polish National Agency for Academic Exchange under grant
No. PPN/ULM/2019/1/00220 and by the European Union HORIZON2020
research and innovation programme under Grant Agreement No. 654002 - ENSAR2.
We acknowledge support by Spanish MCIN/AEI/ 10.13039/501100011033 under
Grants RTI2018-098868-B-I00 and PID2021-126998OB-I00.


\providecommand{\noopsort}[1]{}\providecommand{\singleletter}[1]{#1}%

\end{document}